\documentclass[twocolumn]{aastex631}

\usepackage{amsmath,amssymb}

\usepackage{subfiles}
\usepackage{xr} 
\usepackage{CJK}

\graphicspath{{./}{figures/}}

\begin{document}
\begin{CJK*}{UTF8}{gbsn}

\title{Magnetic field intermittency in the solar wind: PSP and SolO observations ranging from the Alfv\'en region out to 1 AU}

\author[0000-0002-1128-9685]{Nikos Sioulas}
\affiliation{Earth, Planetary, and Space Sciences, 
University of California, Los Angeles \\
Los Angeles, CA 90095, USA }

\author[0000-0001-9570-5975]{Zesen Huang (黄泽森)}
\affiliation{Earth, Planetary, and Space Sciences, 
University of California, Los Angeles \\
Los Angeles, CA 90095, USA }

\author[0000-0002-2381-3106]{Marco Velli}
\affiliation{Earth, Planetary, and Space Sciences, 
University of California, Los Angeles \\
Los Angeles, CA 90095, USA }

\author[0000-0002-7174-6948]{Rohit Chhiber}
\affiliation{Department of Physics and Astronomy, Bartol Research Institute, University of Delaware \\  Newark, DE 19716, USA}
\affiliation{ NASA Goddard Space Flight Center,  \\ Greenbelt,   MD 20771, USA}

\author[0000-0002-7341-2992]{Manuel E. Cuesta}
\affiliation{Department of Physics and Astronomy, Bartol Research Institute, University of Delaware \\  Newark, DE 19716, USA}

\author[0000-0002-2582-7085]{Chen Shi (时辰)}
\affiliation{Earth, Planetary, and Space Sciences, 
University of California, Los Angeles \\
Los Angeles, CA 90095, USA }

\author[0000-0001-7224-6024]{William H. Matthaeus}
\affiliation{Department of Physics and Astronomy, Bartol Research Institute, University of Delaware \\  Newark, DE 19716, USA}

\author[0000-0002-6962-0959]{Riddhi Bandyopadhyay}
\affiliation{ Department of Astrophysical Sciences, Princeton University, \\
 Princeton, NJ 08544, USA}

\author[0000-0002-8700-4172]{Loukas Vlahos}
\affiliation{Deprtment of Physics,  Aristotle University of Thessaloniki\\
GR-52124 Thessaloniki, Greece }

\author[0000-0002-4625-3332]{ Trevor A. Bowen}
\affiliation{Space Sciences Laboratory, University of California, \\ Berkeley, CA 94720-7450, USA}


\author[0000-0001-8358-0482]{Ramiz A. Qudsi}
\affiliation{Department of Physics and Astronomy, Bartol Research Institute, University of Delaware, \\  Newark, DE 19716, USA}

\author[0000-0002-1989-3596]{Stuart D. Bale }
\affiliation{Space Sciences Laboratory, University of California, \\ Berkeley, CA 94720-7450, USA}
\affiliation{Physics Department, University of California, \\ Berkeley, CA 94720-7300, USA}

\author[0000-0002-5982-4667]{Christopher J. Owen}
\affiliation{Mullard Space Science Laboratory, University College London, Dorking, RH5 6NT, UK}

\author[0000-0003-2783-0808]{P. Louarn}
\affiliation{
IRAP, Université Toulouse III - Paul Sabatier, CNRS, CNES, Toulouse, France}

\author[0000-0002-9975-0148]{A. Fedorov}
\affiliation{
IRAP, Université Toulouse III - Paul Sabatier, CNRS, CNES, Toulouse, France}

\author[0000-0001-6172-5062]{Milan Maksimovi\'c}
\affiliation{D\'epartement de Recherche Spatiale et Unit\'e de Recherche associ\'ee au CNRS 264, Observatoire de Paris, Section de Meudon, F-92195 Meudon Principal Cedex, France}

\author[0000-0002-7728-0085]{Michael L. Stevens}
\affiliation{Harvard-Smithsonian Center for Astrophysics, \\ Cambridge, MA 02138, USA}

\author[0000-0002-7077-930X]{Justin Kasper}
\affiliation{BWX Technologies, Inc., Washington DC 20002, USA}
\affiliation{Climate and Space Sciences and Engineering, University of Michigan, Ann Arbor, MI 48109, USA}

\author[0000-0001-5030-6030]{Davin Larson}
\affiliation{Space Sciences Laboratory, University of California, Berkeley, CA 94720-7450, USA}

\author[0000-0002-0396-0547]{Roberto Livi}
\affiliation{Space Sciences Laboratory, University of California, Berkeley, CA 94720-7450, USA}


\begin{abstract}


%
 

$PSP$ and $SolO$ data are utilized to investigate  magnetic field intermittency in the solar wind (SW). Small-scale intermittency  $(20-100d_{i})$ is observed to radially strengthen when methods relying on higher-order moments are considered ($SF_q$, $SDK$), but no clear trend is observed at larger scales. However, lower-order moment-based methods (e.g., PVI) are deemed more appropriate for examining the evolution of the bulk of Coherent Structures (CSs), $PVI \ge 3$. Using PVI, we observe a scale-dependent evolution in the fraction of the dataset occupied by CSs, $f_{PVI \ge 3}$. Specifically, regardless of the SW speed, a subtle increase is found in $f_{PVI\ge3}$ for $\ell =20 d_i$, in contrast to a more pronounced radial increase in CSs observed at larger scales. Intermittency is investigated in relation to plasma parameters. Though, slower SW speed intervals exhibit higher $f_{PVI \geq 6}$ and higher kurtosis maxima, no statistical differences are observed for $f_{PVI \geq 3}$. Highly Alfv\'enic intervals, display lower levels of intermittency. The anisotropy with respect to the angle between the magnetic field and SW flow, $\Theta_{VB}$ is investigated. Intermittency is weaker at $\Theta_{VB} \approx 0^{\circ}$ and is strengthened at larger angles. Considering the evolution at a constant alignment angle, a weakening of intermittency is observed with increasing advection time of the SW. Our results indicate that the strengthening of intermittency in the inner heliosphere is driven by the increase in comparatively highly intermittent perpendicular intervals sampled by the probes with increasing distance, an effect related directly to the evolution of the Parker spiral.
\end{abstract}
\keywords{Parker Solar Probe, Solar Orbiter, Solar Wind, MHD Turbulence, Intermittency}

\section{Introduction}\label{sec:intro}



Powered by the internal dynamics of the Sun, the Solar Wind (SW), a weakly collisional stream of charged particles, expands supersonically into the interplanetary medium carrying with it photospheric magnetic field lines, producing a magnetized sphere of hot plasma around the Sun, the heliosphere. \citep{parker_dynamics_1958, velli_supersonic_1994}. Magnetic field and velocity fluctuations resulting from dynamic processes (e.g., magnetic reconnection) in the solar corona or local dynamics (e.g., driven by stream interactions) that extend over a wide range in the frequency domain have been observed in the solar wind \citep{belcher_large-amplitude_1971, bruno_solar_2013}. During the expansion, nonlinear interactions among the fluctuations result in a cascade of energy towards the smaller scales, and a character that resembles the hydrodynamic turbulence emerges \citep{coleman_turbulence_1968, roberts_velocity_1992}. \par 

\cite{kolmogorov_local_1941} (hereafter \textbf{K41}), derived the relationship between scale dependent increment moments, or structure functions for longitudinal velocity increments, at spatial separation  $\ell$, $ \langle \delta u^{q}_{\ell} \rangle \ = \ \langle | \textbf{u}(\textbf{r}+ \boldsymbol\ell\ ) - \textbf{u}(\textbf{r})|^{q} \rangle$, and the global energy dissipation rate  $\epsilon$ 
\begin{equation} \label{eq:1}
    S_{q} \ = \ \langle \delta u^{q}_{\ell} \rangle \sim \epsilon^{q/3} \ell^{\zeta_{q}}.
\end{equation} \label{eq:structure_functions_intro}

For a full derivation see \citep{rose_fully_1978, frisch_turbulence_1995}. In Eq. \ref{eq:1}, the scaling of field increments occurs with a unique exponent, $\zeta_{q} = q/3$, implying global scale invariance (self-similarity) of the fluctuations and a transfer energy rate that is independent of scale. The consequence of energy conservation in the inertial range can then also allow us to derive the relationship describing the distribution of energy among spatial scales in  Fourier space as $E(k) \sim k^{-5/3}$, corresponding to $\zeta_{2} = 2/3$. Similarly, in the solar wind, a wealth of information about turbulence can be obtained by studying the second statistical moment of the probability distribution function of the magnetic field fluctuations, or Power Spectral Density (PSD). Adopting Taylor's frozen-in hypothesis \citep{taylor_spectrum_1938}, 

\begin{equation}\label{eq:Taylor}
  \kappa  = \frac{2 \pi f_{sc}}{V_{SW}} ,
 \end{equation} 
  
  where, $V_{SW} $ is the solar wind speed, the measured power spectral density in the frequency domain $F(f_{sc})$, is equivalent to the wavenumber power-spectrum $E(\kappa)$ through

\begin{equation}
   E(\kappa) = \frac{V_{sw}}{2 \pi } F(f_{sc}) 
\end{equation}

Due to the nature of the physical processes taking place at different scales, the magnetic spectrum of the solar wind can be divided into several segments, each showing a power-law dependence over wavenumber, $E(\kappa) \propto \kappa^{-\gamma}$.
At the largest scales, a spectral break separates the inertial from the injection scales, where the spectrum is characterized by a $\kappa^{-1}$ dependence. In the fast solar wind, the break shifts to larger scales with increasing heliocentric distance \citep{bruno_solar_2013}. At the intermediate scales, the spectrum steepens with the index, $\gamma$, taking values in the range 3/2 to 5/3 \citep{bavassano_radial_1982, marsch_spectral_1990, matthaeus_measurement_1982, chen_evolution_2020, shi_alfvenic_2021, telloni_evolution_2021}. In the inertial range, fluctuations are described within the simplified nonlinear, incompressible magnetohydrodynamic (MHD) framework, and the energy that was injected into the system at the largest scales, probably of coronal origin, cascades via nonlinear, energy-conserving interactions among the oscillating modes to smaller scales \citep{matthaeus_who_2011}. Once the cascade reaches ion scales, plasma dynamics is governed by kinetic processes, the spectrum steepens, and turbulent energy is converted to plasma heat through mechanisms such as ion cyclotron damping, kinetic Alfv\'en waves, kinetic scale current sheets, etc. \citep{dmitruk_test_2004, tenbarge_current_2013, karimabadi_coherent_2013}. The turbulent cascade is considered to be one of the main processes contributing to the non-adiabatic expansion, as well as the acceleration of the SW \citep[see][and references therein]{matthaeus_who_2011}. Thus, for an accurate description of the dynamics of the heliospheric plasma, understanding the origin and evolution of MHD turbulence is crucial \citep{bruno_solar_2013}. However, even though the analysis of conventional spectral properties can be informative, the second statistical moment of the probability distribution function of increments is only sufficient to fully characterize turbulence under the assumption of isotropic and scale-invariant fluctuations \citep{kolmogorov_refinement_1962}. In practice, these conditions are in principle violated in space and astrophysical systems. As a matter of fact, departures from the linear scaling prediction in Eq. \ref{eq:1}, have been observed for $q$ greater than 3  (e.g., \cite{burlaga_intermittent_1991, carbone_arbitrary-order_2018, chhiber_subproton}), giving rise to the concept of intermittency in solar wind turbulence. These results indicate that to fully comprehend the statistics of turbulent fluctuations, the study of higher-order moments of the scale-dependent probability distribution function of increments is necessary \citep{frisch_turbulence_1995}. 

To better understand intermittency, we may model the turbulent cascade as an effort of the system to approach thermal equilibrium \citep{matthaeus_intermittency_2015}. At shorter time scales, local turbulent relaxation may occur, giving rise to local correlations in MHD. The borders of such regions will typically not be relaxed but rather remain in a dynamic state, leading to local nonlinear interactions and processes such as magnetic reconnection or various types of instabilities. These boundaries correspond to coherent structures $(CSs)$ , which in the case of the solar wind, can be either of coronal origin being passively advected by the SW \citep{borovsky_solar-wind_2021}, or generated locally as an intrinsic feature of the ongoing nonlinear turbulent relaxation process  \citep{matthaeus_selective_1980, veltri_mhd_1999, greco_stochastic_2010, matthaeus_who_2011, matthaeus_intermittency_2015}. Intermittency is associated with a fractally distributed population of small-scale CSs, superposed on a background of random fluctuations \citep{isliker_particle_2019, chhiber_clustering_2020, sioulas_stochastic_2020}. Even though they represent a minor fraction of the entire dataset \citep{osman_intermittency_2012, sioulas_statistical_2022}, CSs account for a disproportionate amount of magnetic energy dissipation, and have been shown to strongly influence the heating and acceleration of charged particles \citep{karimabadi_coherent_2013, osman_intermittency_2012, tessein_association_2013, bandyopadhyay_observations_2020, qudsi_observations_2020, Lemoine_accel_MHD_turb, sioulas_particle_2022}. The fundamental approach to studying intermittency involves the examination of the probability density functions (PDFs) of the dissipation rate. However, based on the Kolmogorov refined similarity hypothesis \textbf{KRSH} \citep{kolmogorov_refinement_1962}, local averages of dissipation rate are related to the scale-dependent increments of the velocity field. Thus, intermittency is reflected on the PDF of field increments in the form of an increasing divergence with respect to a Gaussian distribution (i.e., PDFs display fatter tails) as increasingly smaller scales are involved \citep{castaing_velocity_1990, frisch_turbulence_1995}. This behavior, often referred to as multifractal, violates the concept of global scale invariance, a key assumption of the K41 theory, giving rise to the concept of local scale invariance, i.e., turbulence is characterized by a diverse set of fractals with varying scalings. 

At the same time, solar wind is an expanding medium. MHD fluctuations entering the super-Alfv\'enic wind in the trans-Alfv\'enic region, expected at $\sim 15 -25 R_{\odot}$ \citep{deforest_highly_2018}, are modified in terms of structure and scale as the SW expands into the interplanetary medium driven by the turbulent cascade, as well as, shear at stream interfaces \citep{roberts_velocity_1992} and other transients \citep{shi_influence_2022}. It is, therefore, reasonable to expect that the statistical signatures of coherent structures evolve with heliocentric distance. Indeed, recent studies in the solar wind suggest a dynamic evolution of intermittency properties of MHD fluctuations indicating that the solar wind is an active turbulent medium involving both local and global dynamical processes that influence the higher-order statistics of fluctuations. \cite{bruno_radial_2003} have utilized Helios data to examine the radial evolution of intermittency utilizing the flatness (i.e., SDK  hereafter) of the magnetic field. Their analysis indicates a different behavior for slow and fast wind intermittency. More specifically, slow wind ($V_{SW} \lesssim 500 \  km \cdot s^{-1}$) was observed to display a higher degree of intermittency than the fast wind ($V_{SW} \gtrsim 600 \  km \cdot s^{-1}$). Additionally, no radial dependence was observed for the slow wind, in contrast to an increase in intermittency with heliocentric distance for the fast solar wind. The distinct nature and radial evolution of intermittency were attributed to the different roles played by coherent non-propagating structures and by stochastic Alfv\'enic fluctuations for the two types of wind at different heliocentric distances. Turbulence in fast streams closer to the Sun is highly Alfv\'enic  (i.e., the magnetic field and velocity fluctuations exhibit a high degree of correlation) and displays a self-similar (i.e., monofractal) character. However, during the expansion, due to nonlinear interaction amongst counter-propagating Alfv\'en waves, the fluctuations become decorrelated \citep{roberts_velocity_1992,  chen_evolution_2020, shi_alfvenic_2021} and the Alfv\'enic contribution, which tends to decrease intermittency because of its stochastic nature, is gradually depleted \citep{Marsch_liu_1993}. On the contrary, advected structures tend to increase intermittency because of their coherent nature, while their relative contribution becomes more important with increasing heliocentric distance. As a result, the fractal nature of the magnetic field is modified, gradually approaching  multifractal with increasing heliocentric distance. Slow wind does not show the same behavior since Alfv\'enic fluctuations have a less dominant role for this type of wind. The same line of reasoning was adopted by \citep{ alberti_scaling_2020, telloni_evolution_2021} to interpret the increasing
deviation, with respect to the linear scaling expected from K41 theory, of the structure function scaling exponents (see Equation \ref{eq:structure_functions_intro}), as well as, \cite{greco_evidence_2012} who, utilizing the PVI method, observed an increase in the fractional volume occupied by coherent in the inner heliosphere.
More recently, \citep{Parashar_2019, cuesta_intermittency_2022} have examined the relationship between $SDK$ and $R_{e}$, where $R_e$ is the Effective Reynolds Number, to show that regions with lower $R_{e}$ have on average lower kurtosis, at a fixed physical scale, suggestive of a less intermittent behavior. Even though $Re$ is observed to decrease in the inner heliosphere, several effects  overcome the relation of $Re$  with intermittency, but at 1 au, a change of system dynamics begin to favor the effects from system size, resulting in progressively weaker intermittency at larger heliocentric distances, concurrent with a decreasing $R_{e}$. \par

During its first ten encounters with the Sun, the Parker Solar Probe ($PSP$) mission \citep{fox_solar_2016}
has provided valuable measurements of solar wind particles and fields in the neighborhood of the solar wind sources. Aiming to approach the surface of the Sun by as close as $9.86 \ R_{\odot}$, PSP offers unprecedented in-situ measurements in the vicinity of the Alfv\'en-zone, allowing us to study its influence on the evolution of spectral and intermittency properties of the field fluctuations \citep{kasper_PSP_alfven, Bandyopadhyay_alfven, Zhao_alfven_region}. These observations will supplement  simulations \citep{chhiber2022MNRAS} and ultimately enable us to explore processes such as the heating of the solar corona and the acceleration of the solar wind in the vicinity of the Alfv\'en zone\citep{matthaeus_who_2011}. In conjunction with the recently launched Solar Orbiter $SolO$ \citep{muller_solar_2020}, the synergy of the two missions offers a unique opportunity to explore the connection between the Sun and the heliosphere. \par

In this paper, we are interested in understanding the radial evolution of inertial range MHD turbulence and studying the basic features of scaling laws for solar wind fluctuations. We start our investigation by examining the radial evolution of intermittency without accounting for the anisotropy introduced with respect to the alignment angle, $\Theta_{VB}$. At a later stage, however, we show that accounting for anisotropy will complicate interpretation of the observations.

For the purposes of our analysis, high-resolution magnetic field and particle data from PSP and SolO covering heliocentric distances $13\lesssim R\lesssim 220$~$R_{\odot}$ are implemented.
Our tools to study intermittency involve analytical methods such as the Partial Variance of increments (PVI), the scaling behavior of the high order moments of variations of the magnetic fields separated by a scale $\ell$, or Structure Functions ($SFs$), and their respective scaling exponents, and finally the Scale Dependent Kurtosis ($SDK$) of the magnetic field.

The structure of this paper is as follows: Section \ref{sec:higher order statistics and intermittency} introduces the diagnostics of intermittency utilized in this study; Section \ref{sec:methods} presents the selection of data (PSP, Solar Orbiter) and their processing; The results of this study are presented in Section \ref{sec:results}: In Subsection \ref{subsec:radial_evolution_intermittency}, the radial evolution of magnetic field intermittency is investigated, and in Subsection \ref{subsec:intermittency_plasma_params} the dependence of intermittency on plasma parameters is examined; A summary of the results along with the conclusions is given in Section \ref{sec:Summary}.

\section{Higher-order statistics and intermittency}\label{sec:higher order statistics and intermittency}

\subsection{Radial evolution of intermittency in the Solar Wind. The importance of normalizing spatial scales.}\label{subsec:Why_normalize}

\begin{figure}
\centering

\includegraphics[width=0.4\textwidth]{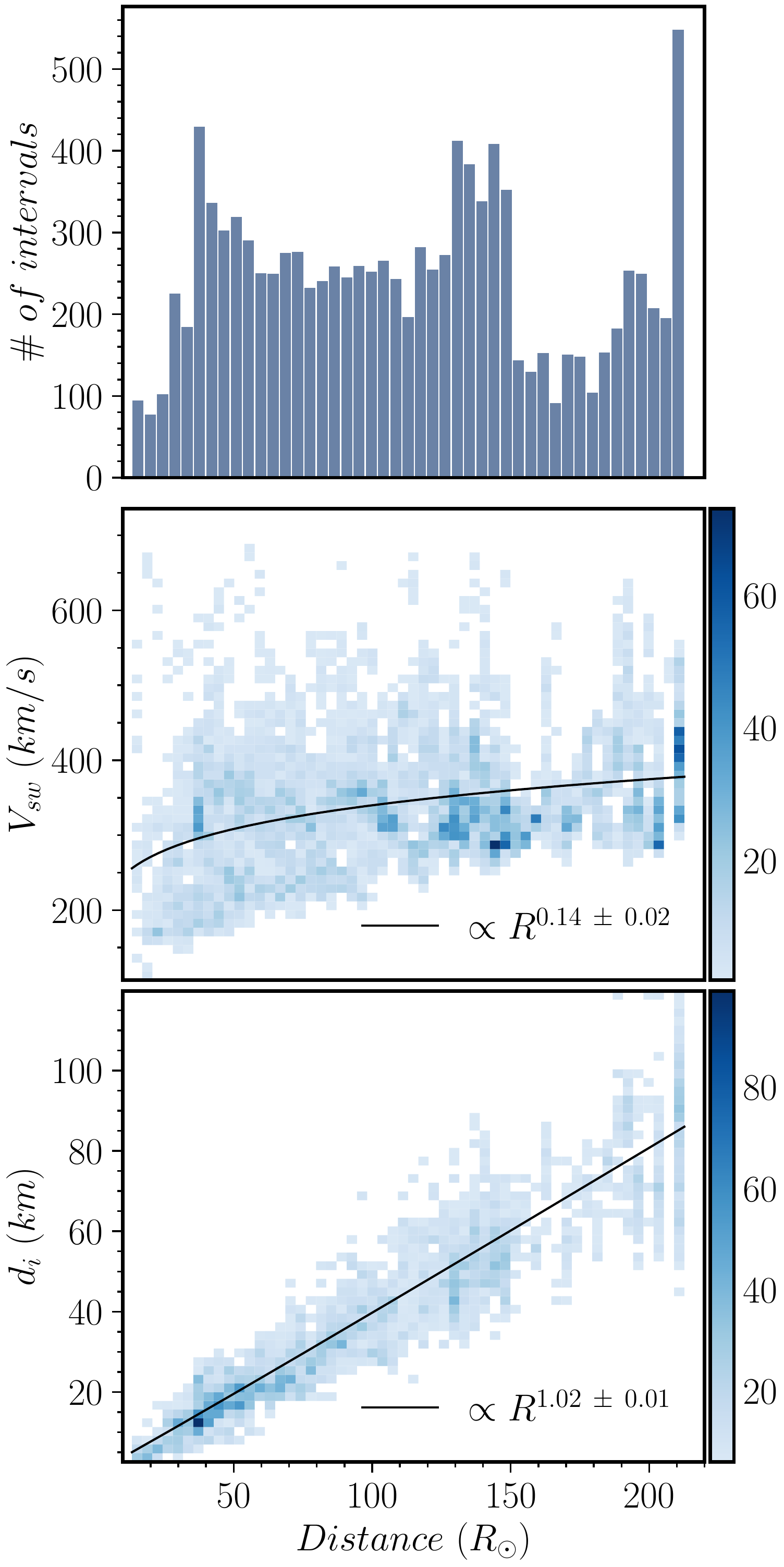}
\caption{ (a) Histogram showing the number of 5-H intervals as a function of heliocentric distance  (b) Joint distribution of SW speed $V_{sw}$ vs heliocentric distance (R) (c) Joint distribution of ion inertial length $d_{i}$ vs heliocentric distance(R) .}
\label{fig:relev_quant1}

\end{figure}

When studying the radial evolution of intermittency in the solar wind, one fundamental issue arises: Due to the expansion of the solar wind, ion inertial length $d_i$ is expected to increase linearly as a function of heliocentric distance \citep{cuesta_intermittency_2022}. 
At the same time, PDFs of normalized magnetic field increments have been shown to display fatter tails when smaller spatial scales are considered \citep{bruno_solar_1999}. This indicates that when the radial evolution of higher-order statistics is investigated, the use of a constant  (i.e., not normalized) lag will result in calculating the intermittency diagnostics on progressively smaller spatial scales. It is thus becoming obvious that to reveal the underlying physical processes on a constant scale, it is important to normalize the spatial scales with physically relevant plasma parameters. For this reason, temporal scales are first converted to spatial scales by means of Taylor's Hypothesis \citep{taylor_spectrum_1938}
\begin{equation}\label{eq:Taylor}
  \ell  = -V_{SW}  \tau,
\end{equation}
where $V_{SW}$ is the solar wind speed, and $\tau$ is the temporal lag. Taylor's hypothesis is founded on the idea that speeds of MHD wave modes (e.g., shear Alfv\'en modes, propagating at $V_p=V_A\cos(<\boldsymbol k, \boldsymbol B>)$, where $V_{A} = B / \sqrt{\mu_{0} \rho}$ is Alfv\'en speed) observed in  the solar wind plasma are insignificant compared to the bulk flow of the solar wind (i.e., the Alfv\'en Mach Number $M_{A} = \frac{V_r}{V_A} \ll 1$). Indeed, with the exception of a few sub-Alfv\'enic intervals during $E_{8}-E_{11}$ of PSP, for the vast majority of the intervals, $M_{A} < 1$ was satisfied. Note, however, that even when $M_{A} \sim 1$ the Taylor hypothesis may still be applied \citep{perez_how_2021}. Subsequently, the ion inertial length $d_i$ is utilized to normalize the spatial scales. The ion inertial length is defined as
\begin{equation}
  d_{i} = \frac{c}{\omega_{pi}}=228 \cdot \tilde{n}_{p}^{-1/2}  \ [km],
  \label{eq:di}
\end{equation}
where $c$ is the speed of light and $\omega_{pi}$ is the proton plasma frequency, and $\tilde{n}_{p}$ is the mean proton density for the respective interval. By means of integrating the conservation laws over a spherically expanding surface, a spatial scaling of $n_{p} \propto R^{-2}$ is expected. Taking into account Equation \ref{eq:di}, the ion inertial length is then expected to scale with distance as $d_{i} \propto R$. In Figure \ref{fig:relev_quant1}, the observed scaling of $d_i$ is presented, consistent with the theoretical expectation, as well as with previous Helios and Voyager observations \citep{cuesta_intermittency_2022}. The number of 5 hr-long intervals analyzed as a function of distance as well as the advection time of the solar wind is also presented in the same figure. \\

\subsection{Partial Variance of Increments}\label{subsec:Results_PVI}

 The boundaries of CSs are associated with spatial variations or reversals of the local magnetic field. In recent years, a variety of methods, suitable for the detection of sharp gradients in a turbulent field, have been proposed \citep{bruno_solar_1999, hada_phase_2003, khabarova_automated_2021, pecora_f_identification_2021}. A convenient statistical tool to perform this study, is the Partial Variance of Increments $PVI$ \citep{greco_intermittent_2008}. The PVI method has been used in the past in a variety of space plasma environments to determine the portion of the data corresponding to the underlying CSs \citep{tessein_association_2013, chasapis_thin_2015, bandyopadhyay_observations_2020, chhiber_clustering_2020,Vasko_CS_21,Lotekar_Vasko_21}. Assuming the validity of Taylor's hypothesis \citep{taylor_spectrum_1938}, the $PVI$ index at time $t$, for lag $\ell = - V_{SW} \tau$, is given by \citep{greco_intermittent_2008}:

\begin{equation}\label{eq:PVI}
    PVI( t , \ell) \ = \  \frac{| \delta\boldsymbol{B}( t, \tau)|}{\sqrt{\langle | \delta\boldsymbol{B}( t,\tau)|^{2}\rangle}},
\end{equation}

where, $| \delta\boldsymbol{B}( t, \tau)| \ = \ |\boldsymbol{B}(t \ + \ \tau) \ -  \ \boldsymbol{B}(t)| $ is the magnitude of the magnetic field vector increments, and $\langle...\rangle$ stands for an average over a suitably large window that is a multiple of the estimated correlation time for the magnetic field. \cite{greco_partial_2018}  have shown that as the PVI index increases, the identified events are more likely to be associated with Non-Gaussian structures that lay on the "heavy tails" observed in the PDF of scale-dependent increments, suggesting that coherent structures correspond to events of index  $ PVI \geq 2.5$. By further increasing the threshold value $\theta$, one can then identify the most intense magnetic field discontinuities like current sheets and reconnection sites \citep{servidio_magnetic_2009}.

\subsection{Structure Functions and Scaling Exponents }\label{subsec:Theory_SF}

Another method for assessing intermittency in a time series is based on estimating a sequence of $q^{th}$ order moments of the magnetic field increments. The physical significance of this method
is founded on the susceptibility of higher-order moments
to concentrations of energy dissipation related to coherent structures and, from KRSH to extreme values of
the magnetic field increments. We can estimate the component increments of the magnetic field at time $t$ as

\begin{equation}
    \delta B_{i}( t, \tau)  =  B_{i}(t +  \tau) -   B_{i}(t),
\end{equation}

where, $i = {X,Y,Z}$. Temporal lags may be recast to spatial lags  by means of Taylor's approximation, and the $q^{th}$  order structure-function for the magnetic field can be estimated through

\begin{equation}
    S^{q}_{B}(\ell) = \langle [\delta \boldsymbol{B}(t,\ell)]^{q}\rangle_{T},
\end{equation} \label{eq:Structure_functions}

where, $|\delta \boldsymbol{B}(t,\ell)| =  ( \sum_{i}\delta B_{i}^{2} )^{1/2}$, is the magnitude of the vector  magnetic field increments, and $\langle ... \rangle_{T}$ stands for averaging over an interval of duration  T. For increment scale $\ell$, the moments are expected to display a power-law dependence, $S^{q}_{B}(\ell)  \propto \ell^{\zeta(q)}$. Thus, after the moments are calculated, power-law fits may be applied on the curves over different ranges of spatial scales to obtain the scaling exponents $\zeta_{q}$ of the structure functions. Based upon the assumptions that the statistical properties of the turbulent fields are locally homogeneous and isotropic, i.e., the energy dissipation rate within the inertial range is constant on average, both the K41 theory \citep{kolmogorov_local_1941} in hydrodynamics, as well as, the Iroshnikov-Kraichnan \citep{iroshnikov_turbulence_1963} model for MHD turbulence predict a linear scaling of $\zeta(q)$ with order q in the fully developed regime, where $\zeta_{q} =q/3$, and $\zeta_{q} =q/4$ respectively. However, as denoted by Landau \citep{ oboukhov_1962, kolmogorov_refinement_1962}, irregularity of energy dissipation is expected to alter the scaling exponents of field increments with $\ell$. More specifically, in the case where $\epsilon$ statistically depends on scale due to the mechanism that transfers energy from larger to smaller eddies, $\epsilon$ should be replaced by $\epsilon_{\ell}$  and Equation \ref{eq:structure_functions_intro} should be recast to 
\begin{equation}
    S^{q}_{B} \sim \langle \epsilon_{\ell}^{q/3}\rangle \ell^{q/3}.
\end{equation}
 Expressing $\epsilon_{\ell}^{q/3}$ via a scaling relation with $\ell$ we obtain 
\begin{equation}
    \langle \epsilon_{\ell}^{q/3}\rangle  \sim  \ell^{\tau_{q}/3}
\end{equation}
and thus 
\begin{equation}
    S^{q}_{B} \sim  \ell^{\zeta(q)},
\end{equation} 
where, $\zeta(q)= q/3 + \tau_{q}/3$ is generally a nonlinear function of q. Thus, when the scaling exponents $\zeta_q$ of the structure functions are considered, the different local subsets of fractal dimension within a turbulent field are reflected on the departures from the linear scaling. This departure implies the violation of global scale invariance which in turn indicates a process characterized by multifractal statistics and intermittency.

In recent years, it has been observed that for turbulence that is either not fully developed or is in a system of finite size, one does not have direct access to the scaling exponents $\zeta(q)$. In such cases, the extended self-similarity (ESS) hypothesis \citep{benzi_ESS} posits that it is still possible to investigate the functional form by plotting  structure functions of different order against each other. This implies that the scaling of structure functions of order q  may relate better to the behavior of another structure-function of order p than to spatial lag $\ell$. As a result, ESS establishes the following scaling for the structure functions
\begin{equation}
    S^{q}(\ell) = S^{p}(\ell)^{\zeta(q)/\zeta(p)}.
\end{equation}\label{eq:ESS}

\subsection{Scale Dependent Kurtosis }\label{subsec:Theory_SDK}
An intermittency-affected generic time series exhibits alternate intervals of very high activity followed by extended periods of quiescence. Thus, intermittency in a signal is manifested in the form of a decrease in the fraction of volume occupied by structures at scale $\ell$ with decreasing scale. Thus, due to it's relationship to the scale dependent filling fraction $F(\ell)$ for structures through
\begin{equation}
    K(\ell) \sim 1/F(\ell)
\end{equation}
 the Scale Dependent Kurtosis (SDK), defined as 
\begin{equation}
   \mathcal{K}(\ell) =  \frac{\langle| \delta \boldsymbol{B}|^{4} \rangle}{\langle |\delta \boldsymbol{B}|^{2} \rangle^{2}},
   \label{eqn:SDK Definition}
\end{equation}
 can be utilized to characterize the intermittency of a statistically homogeneous signal \citep{frisch_turbulence_1995}. 
 
 
An increase in  $\mathcal{K}(\ell)$ with the involvement of smaller and smaller scales $\ell$ is indicative of a signal that exhibits activity over only a fraction of space, with the fraction decreasing with the scale $\ell$ under consideration. For a scalar that emerges from an additive random process subject to a central limit theorem (i.e., follows a Gaussian distribution), the SDK is independent of scale and attains a constant value $\mathcal{K}(\ell) = 3$, indicating the self–similar character of the fluctuations. On the contrary, the PDFs of intermittent fluctuations progressively deviate from a Gaussian distribution (i.e., distributions display fat tails) at smaller scales \citep{frisch_turbulence_1995}. As a result, intermittency in a generic timeseries is manifested in the form of a monotonically increasing SDK with the involvement of smaller spatial/temporal scales. Additionally, when comparing two different time series, the one for which SDK grows more rapidly will be considered as more intermittent. Note that in the case where SDK fluctuates around a value different from 3, fluctuations are still characterized as self–similar but not Gaussian (i.e., formally referred to as Super-Gaussian). In this case, the fluctuations are not considered intermittent. This can be better understood when considering the way PDFs of increments are modified with scale. For instance, one may consider increments of a field $\phi$ that follow a given scaling
\begin{equation}
    \delta \phi_{\ell} = \langle |\phi(x + \ell) - \phi(x )| \rangle \sim \ell^{h}.
\end{equation}

By introducing a change of scale, $\ell \rightarrow \kappa \ell$, where $\kappa > 0$, we get the following transformation 
\begin{equation}
    \delta \phi_{\kappa \ell} \sim \kappa^{h}\delta \phi_{\ell}.
\end{equation}
According to this relationship, increments estimated at different scales, are characterized by the same statistical properties \citep{frisch_turbulence_1995} 
 \begin{equation}
     PDF(\delta \phi_{\kappa \ell})= PDF(\kappa^{h}\delta \phi_{\ell}).
 \end{equation}
 
 This means that if  $\kappa$ is unique, 
 the PDFs of the normalized increments (e.g., rescaled by their standard deviations),  $\delta \phi_{\ell}(x) = (\phi(x + \ell) - \phi(x ))/ \langle(\phi(x + \ell) - \phi(x ))^{2} \rangle ^{1/2} $ collapses to a  single PDF highlighting the self-similar (fractal) nature of the fluctuations. On the other hand, intermittency implies multifractality and, as a consequence, an entire range of values for $\kappa$. It is thus reasonable to expect that over the scales for which the PDF of increments collapse on to each other, the SDK will fluctuate around a constant value.


\section{Methods}\label{sec:methods}

\subsection{Data Selection}\label{subsec:Data_selection}

A  merged dataset of PSP and SolO observations is employed to study the radial evolution of magnetic field fluctuations in the inner heliosphere. In the first step, all available PSP observations from the period 2018-11-01 to 2022-02-22 (i.e., orbits 1 - 11 ) are collected. More specifically, Level 2 magnetic field data from the Flux Gate Magnetometer (FGM) \cite{bale_fields_2016}, and Level 3 plasma moments from the Solar Probe Cup (SPC), as well as the Solar Probe Analyzer (SPAN) part of the Solar Wind Electron, Alpha and Proton (SWEAP) suite \citep{kasper_solar_2016}, in the spacecraft frame, have been analyzed. Magnetic field data were obtained at a cadence of 4.58 samples per second ( $\sim 0.218 s$ resolution). However, it was found that the higher cadence is mostly offered close to the perihelia of PSP, while for periods where PSP is further away from the Sun, the cadence is reduced to $~0.42 s$. For plasma moments, the cadence strongly depends on the interval studied, ranging from  $0.218-0.874s$ during the encounters and to $\sim 27.9 s$ at larger heliocentric distances for SPC, while for Span-i, the median cadence is $\sim 3.5$ for the encounters and $\sim 28s$ further away from the sun.

In the second step, magnetic field and particle data from the SolO mission from 2021-01-01 to 2021-12-01 are also employed. Magnetic field measurements from the Magnetometer (MAG) instrument \citep{horbury_solar_2020}, downloaded from the ESA Solar Orbiter archive, have been utilized. Particle moment measurements for our study are provided by the Proton and Alpha Particle Sensor (SWA-PAS) onboard the Solar Wind Analyser (SWA) suite of instruments \citep{owen_solar_2020}.
 
\subsection{Data processing}\label{subsec:Data_processing}

In order to account for gaps in the magnetic field timeseries, the mean value of the cadence between successive measurements $\langle \delta \tau \rangle$ has been estimated for each interval. Subsequently, intervals have been divided into three classes (a) $\langle \delta \tau \rangle \leq 250ms$ (b) $\langle \delta \tau \rangle \leq 500ms$ (c) $\langle \delta \tau \rangle \geq 500ms$. For the first two classes of intervals magnetic field data have been linearly resampled to a cadence of $dt =250 ms$  and $dt =500 ms$ respectively, while the remaining intervals have been discarded. 
 This decision was based on the observation that when resampling to $dt =450 ms$, the minimum spatial scale of $20d_{i}$ could not be achieved for a minor fraction of the intervals at distances $\mathcal{R} \leq 20 \mathcal{R}_{\odot}$. To confirm that the different cadence does not affect the results of our study, the analysis was repeated by resampling all magnetic field data to $dt =450ms$, with qualitatively similar results. 

\begin{figure*}
\centering
\setlength\fboxsep{0pt}
\setlength\fboxrule{0.0pt}
\includegraphics[width=0.45\textwidth]{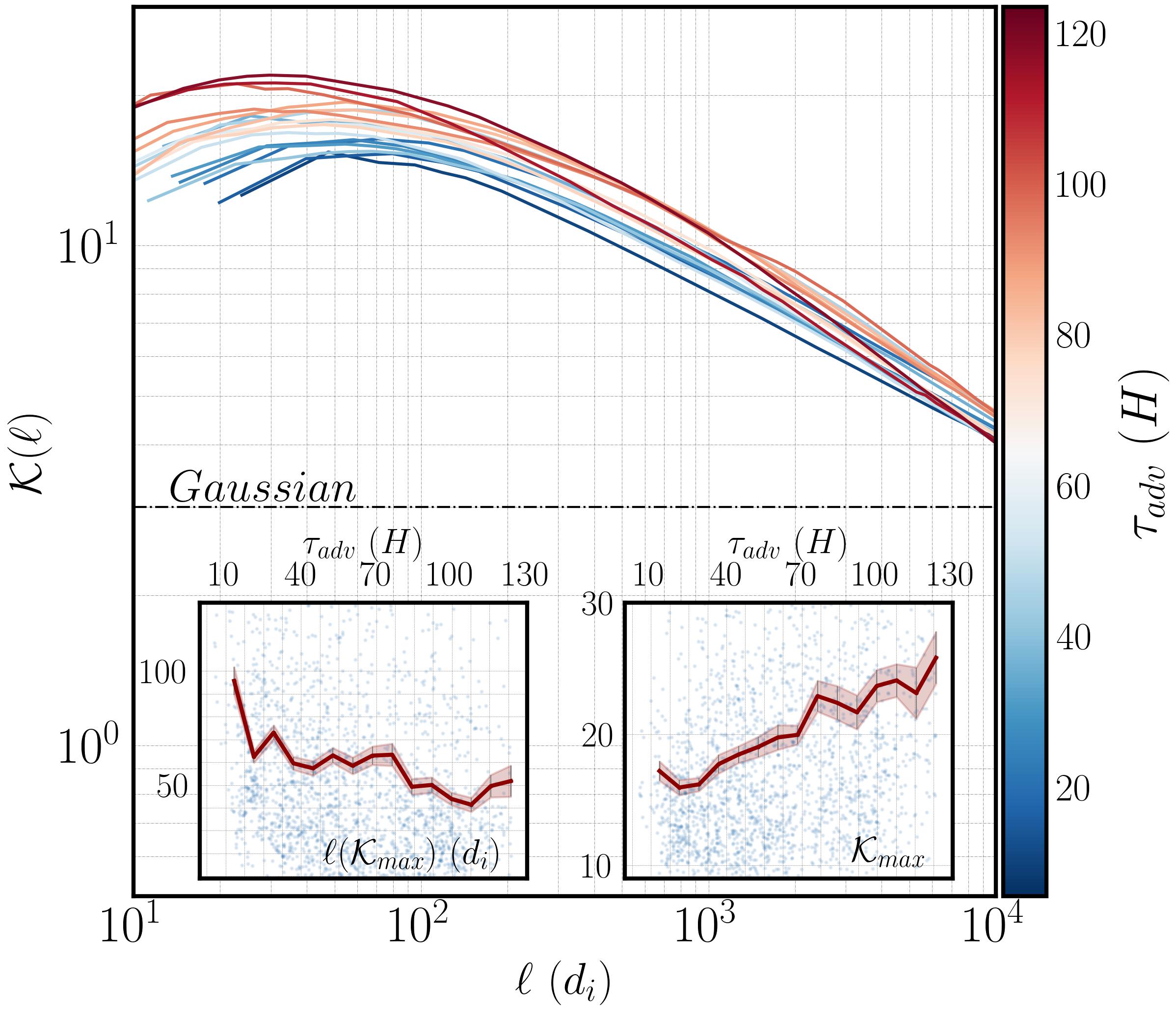}
\includegraphics[width=0.445\textwidth]{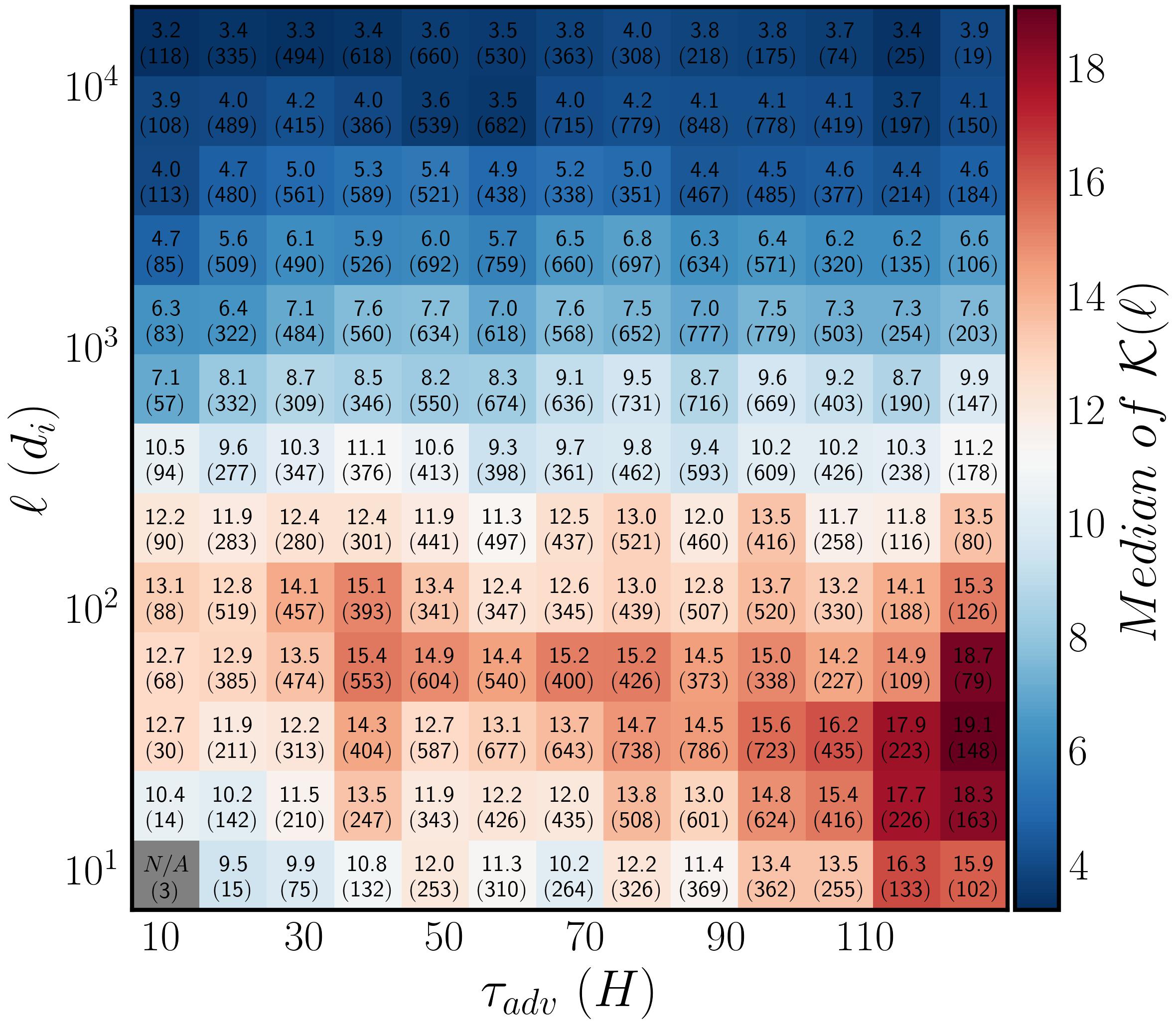}

\caption{(a) Evolution of SDK with $\tau_{adv}$. Each line represents the average of 100 intervals that fall within the same $\tau_{adv}$ bin. The inset scatter plots show (i) the scale, in units of $d_i$,  at which the kurtosis attains the maximum value  (ii) The maximum kurtosis value estimated for the individual 5H intervals. The binned mean of the two quantities is also shown with a red line. (b) Scale-dependent kurtosis of the magnetic field as a function of scale in units of ion inertial length $d_{i}$ and SW convection time $\tau_{adv}$. The numbers indicate the median value of kurtosis within each bin, and the bracketed numbers show the number of events within each bin. }

\label{fig:SDK1}
\end{figure*}


To obtain the plasma parameters from PSP, either SPAN or SPC data are utilized depending on the quality and cadence of the data for the interval. Subsequently, a Hampel filter was applied to all particle moments timeseries to eliminate spurious spikes and outliers exceeding three standard deviations from a moving average window spanning 200 points  \citep{davies_identification_1993}. Finally, in order to maintain a sufficient statistical sample within any given interval, intervals that were found to have more than $5 \%$ of the values missing in the magnetic field or $10 \%$ in the particle timeseries have also been discarded. For the purposes of the SDK and SF's analysis, magnetic field and particle data have been divided into intervals of duration $d= 5 \ hr$. On the other hand, due to the nature of the PVI analysis (i.e., our measure of intermittency is provided in the form of a timeseries) after PVI was estimated, data have been divided into intervals of duration $d = 15 \ min$. The smaller duration intervals were chosen to mitigate the effects of mixing different types of solar wind, as well as to allow us to study intervals for which the alignment angle, $\Theta_{VB}$, estimated as

\begin{equation}
    \Theta_{VB} = cos^{-1}(\frac{\langle \boldsymbol{B}\rangle \cdot \langle \boldsymbol{V_{SW}}\rangle}{\langle ||\boldsymbol{B}||\rangle \cdot \langle ||\boldsymbol{V}_{SW}||\rangle}),
\end{equation}
 where, $||\cdot||$, is a vector magnitude, $\langle \cdot \rangle$ denotes ensemble average, and \textbf{B}, \textbf{V}$_{SW}$ are the magnetic field and solar wind velocity measured in the spacecraft frame, do not vary significantly.
Additionally, in an attempt to interpret the evolution of parameters that measure the evolution of intermittency, we attribute the differing behaviors in various samples of solar wind turbulence to the role played by the advection time of the solar wind defined as $  \tau_{adv} = \frac{D_{SC}}{V_{SW}} \ [Hours],$ where $D_{SC}$ is the heliocentric distance of the spacecraft in units of km and $V_{SW}$ has units of $km \ H^{-1}$. Note that the same analysis was repeated by differentiating the plasma streams based on their heliocentric distance with qualitatively similar results. 

\section{Observational Results}\label{sec:results}

\subsection{Radial evolution of magnetic field intermittency.}\label{subsec:radial_evolution_intermittency}






\subsubsection{Scale Dependent Kurtosis (SDK).}\label{subsubsec:Results_SDK}

In this section,  we investigate the evolution of the magnetic field kurtosis as a function of lag and advection time of the solar wind. Following the methods described in Section \ref{subsec:Theory_SDK}, the SDK for the magnetic field magnitude was estimated for 3100 individual intervals of duration $d =5$ hr. In Figure \ref{fig:SDK1}a the evolution of SDK as a function of advection time $\tau_{adv}$ of the solar wind is presented. To emphasize the trend, the average of 100 intervals that fall within the same $\tau_{adv}$ bin is estimated, and the mean value of $\tau_{adv}$ for each bin is reflected on the line color. At the largest scales, the kurtosis is near Gaussian, and no clear trend is observed. At smaller lags, $\ell \sim 5 \cdot 10^{3} d_{i}$, the lines intersect and beyond this point separate. In particular, for $\ell \lesssim 5 \cdot 10^{3} d_{i}$, an increase in kurtosis is observed with increasing $\tau_{adv}$. For a more quantitative comparison, the evolution of the maximum value of the kurtosis (thereby $K_{max}$) as a function of $\tau_{adv}$ is presented on the right-hand side inset of Figure \ref{fig:SDK1}a. The blue dots indicate $K_{max}$ estimated for individual intervals, and the red line is the binned mean of the same quantity. Uncertainty bars indicate the standard error of the sample, $\sigma_{i}/ \sqrt{n}$, where $\sigma_{i}$ is the standard deviation and $n$ is the number of the samples inside the bin \citep{gurland_simple_1971}. Additionally, the spatial scale $\ell$ at which $K_{max}$ is observed, i.e. $ \ell (K_{max})$, expressed in units of $d_{i}$, is presented in the left inset figure. Even though there is a  considerable scatter in the data, the maximum shifts towards smaller lags with increasing $\tau_{adv}$. Moreover, as $\tau_{adv}$  increases, the peaks of SDK are progressively shifted to larger and larger values. As noted in the introduction, intermittency manifests itself as a growing kurtosis with decreasing spatial scale. Consequently, the increase of $K_{max}$ indicates a radial strengthening in intermittency within the inner heliosphere. Figure \ref{fig:SDK1} offers a different perspective on the evolution of SDK as a function of $\tau_{adv}$ for different spatial scales. The data points were binned according to $\ell$, and $\tau_{adv}$, and the median value inside each bin was calculated, which is reflected in the colors and written in the plot. The bracketed numbers in the plots are the number of data points inside each bin. Note that bins that include less than 10 data points have been discarded. From this figure, we can understand that at small lags, there is a clear upward trend as a function of $\tau_{adv}$, whereas, for larger spatial scales, such trend becomes progressively less obvious. 

\begin{figure}
    \centering
    \includegraphics[width=0.23\textwidth]{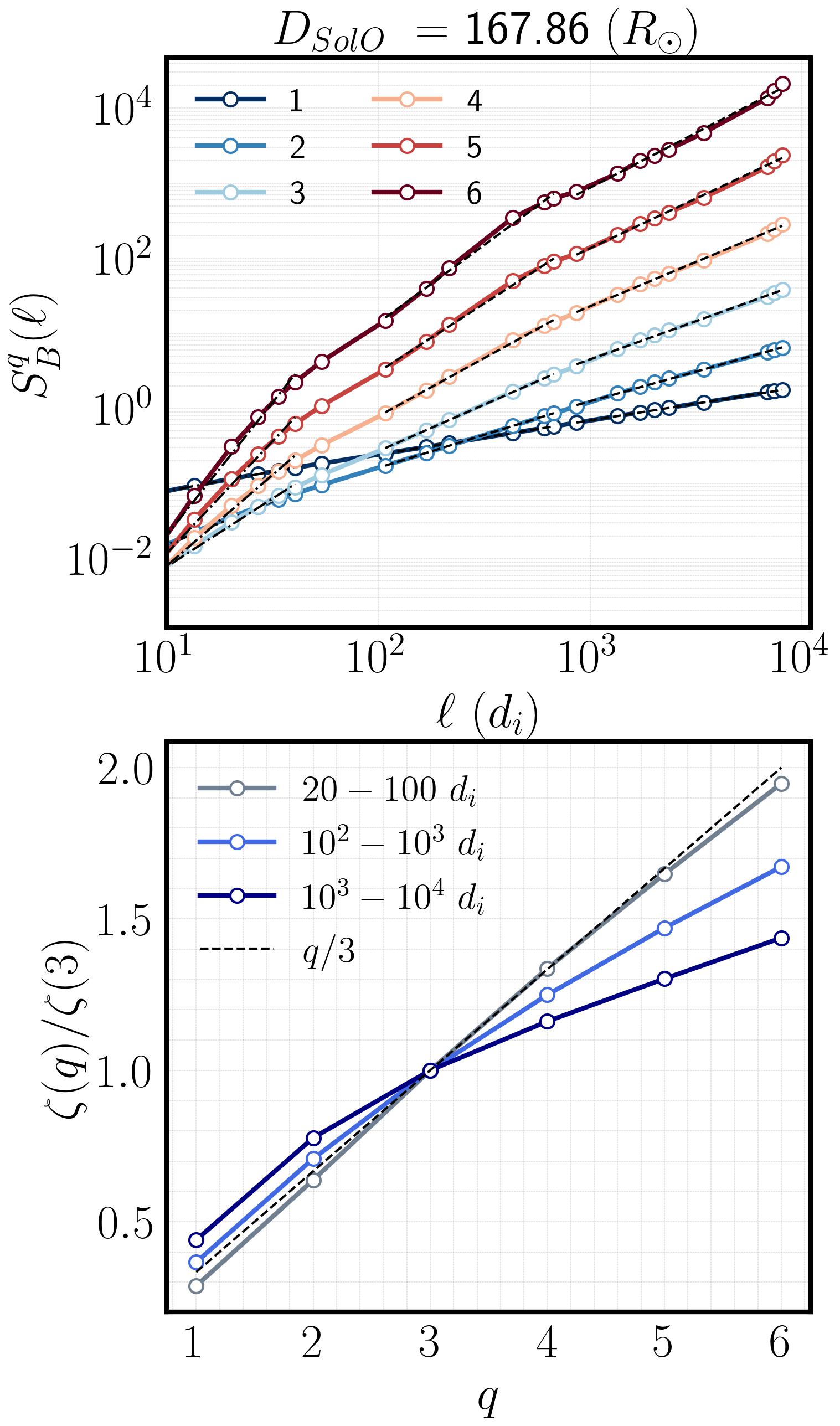}
    \includegraphics[width=0.23\textwidth]{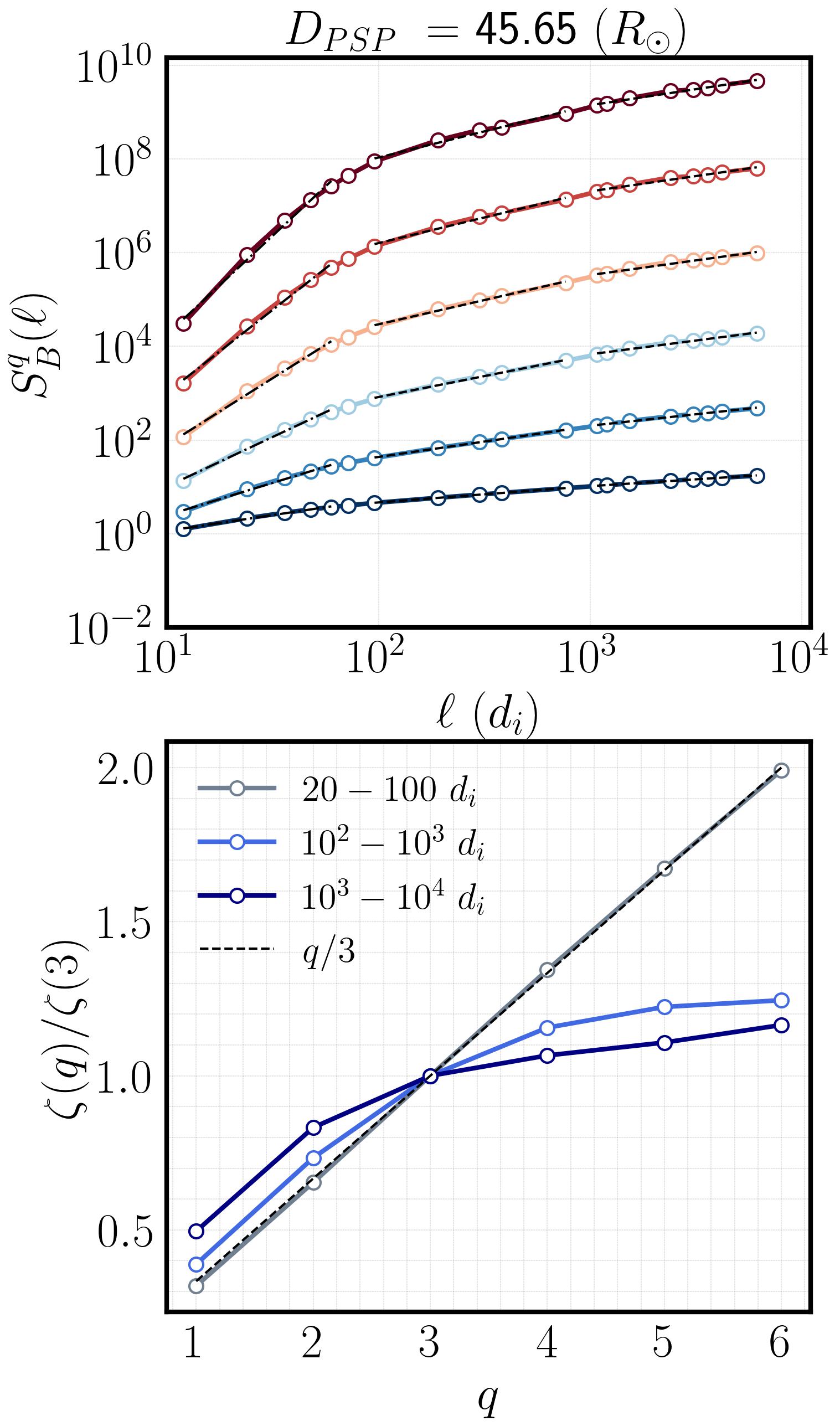}\
\label{fig:sfuncs}

    \caption{(a)  Structure functions  for $\delta \textbf{B}$, as a function of spatial lags in units of the ion inertial length $d_{i}$, $S_{B}^{p}(\ell)$. Power-law fits applied to $S_{B}^{p}(\ell)$, in the range (a)  $20-10^{2}d_{i}$, (b)  $10^{2}-10^{3}d_{i}$ (c) $10^{3}-10^{4}d_{i}$  are shown as a dashed-dotted and dashed line respectively. (b) The normalized resulting scaling exponents $\zeta(q)/\zeta(3)$, estimated from Fig. ref{fig:sfuncs}a  as a function of q. The K41 (q/3) linear scallings is also shown for comparison.}
    \label{fig:SFuncs1}
\end{figure}

\begin{figure*}
\centering
\setlength\fboxsep{0pt}
\setlength\fboxrule{0.0pt}

\fbox{\includegraphics[width=0.329\textwidth]{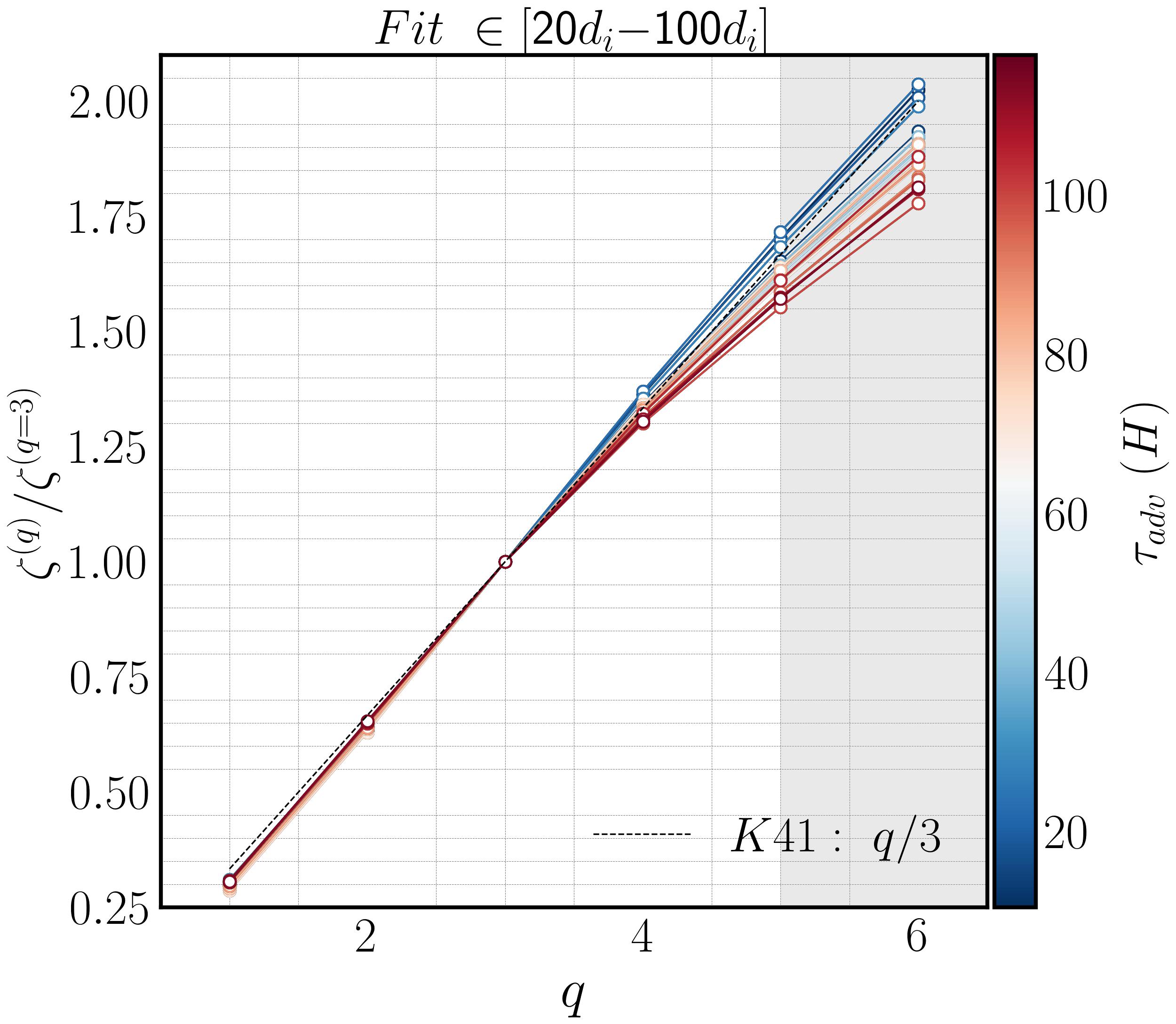}}
\fbox{\includegraphics[width=0.329\textwidth]{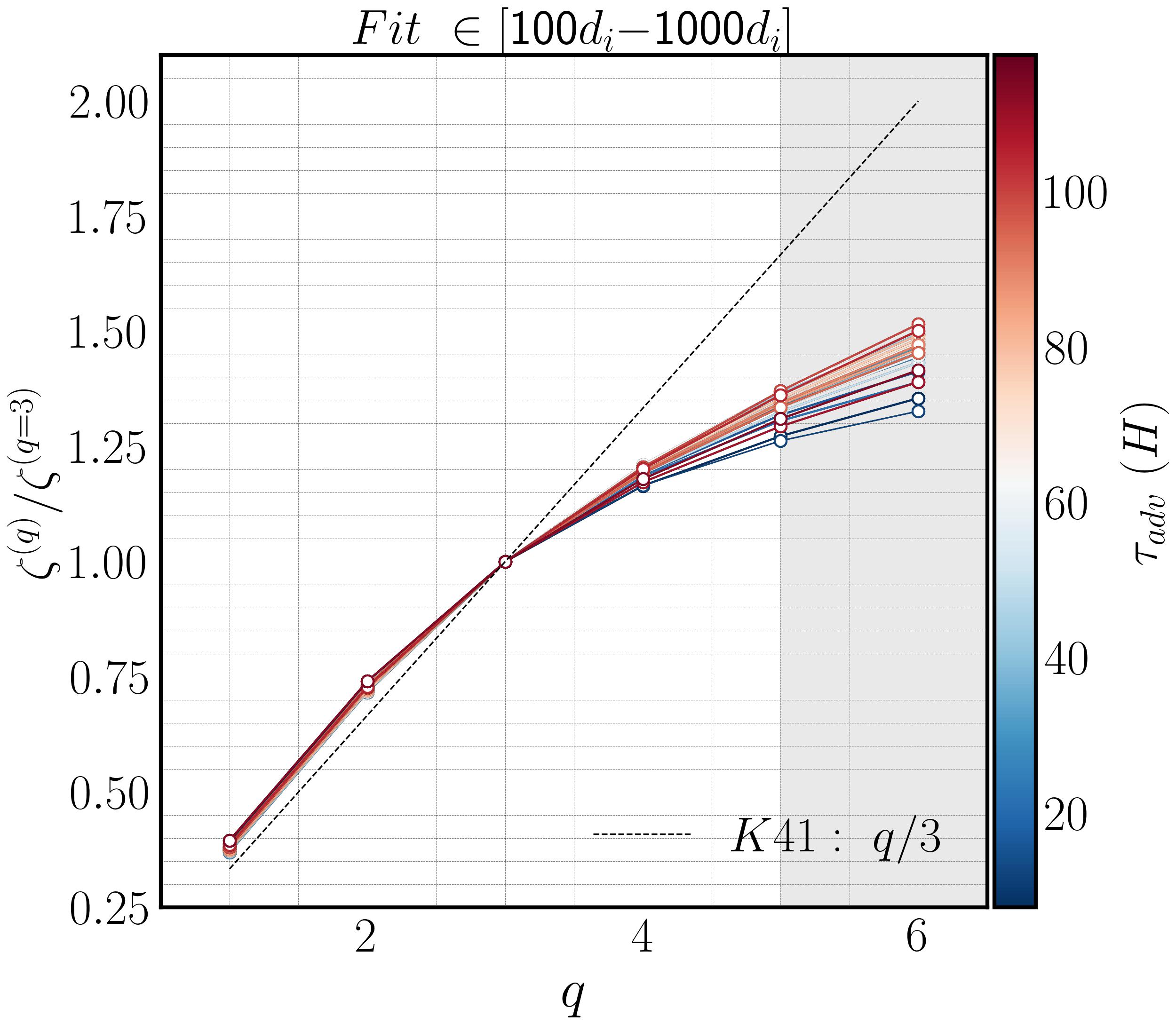}}
\fbox{\includegraphics[width=0.329\textwidth]{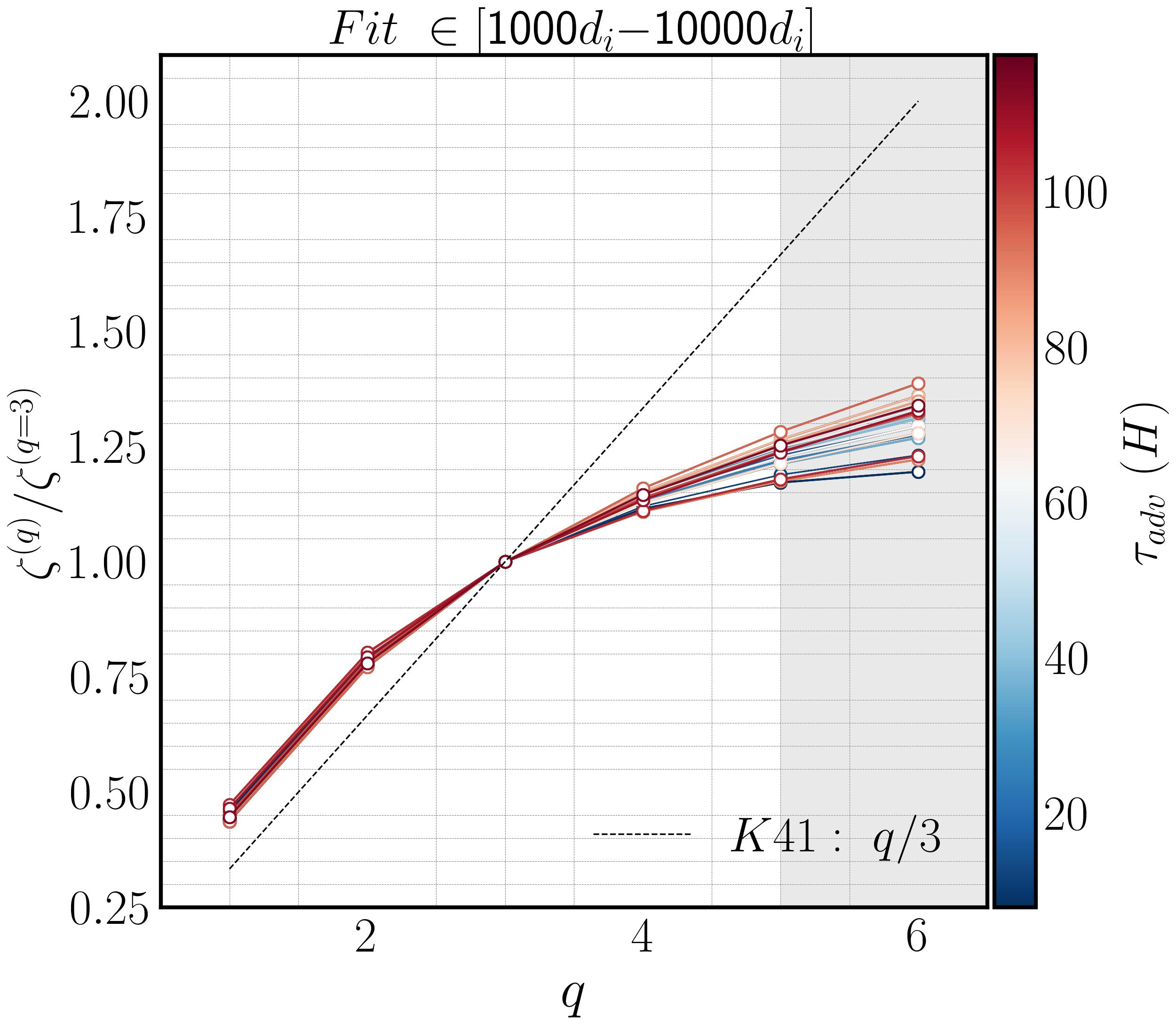}}

\caption{Normalized scaling exponents  $\zeta(q) / \zeta(3)$ vs. $q$, as a function of advection time $\tau_{adv}$. Each line represents the average of 100 intervals that fall within the same $\tau_{adv}$ bin. The scaling exponents have been obtained by applying a power-law fit on the structure functions ($S_{B}^{q}(\ell)$) within three different spatial ranges: (a) $[20 d_i, 100 d_i]$, (b) $[100 d_i, 1000 d_i]$, (c) $[1000 d_i, 10000 d_i]$. The K41 linear scaling is also shown with the black dashed line in all panels. The shaded area has been included to indicate moments that are not determined with reliable accuracy.}
\label{fig:SFuncs2}
\end{figure*}

    

\begin{figure}
\centering
\includegraphics[width=0.38\textwidth]{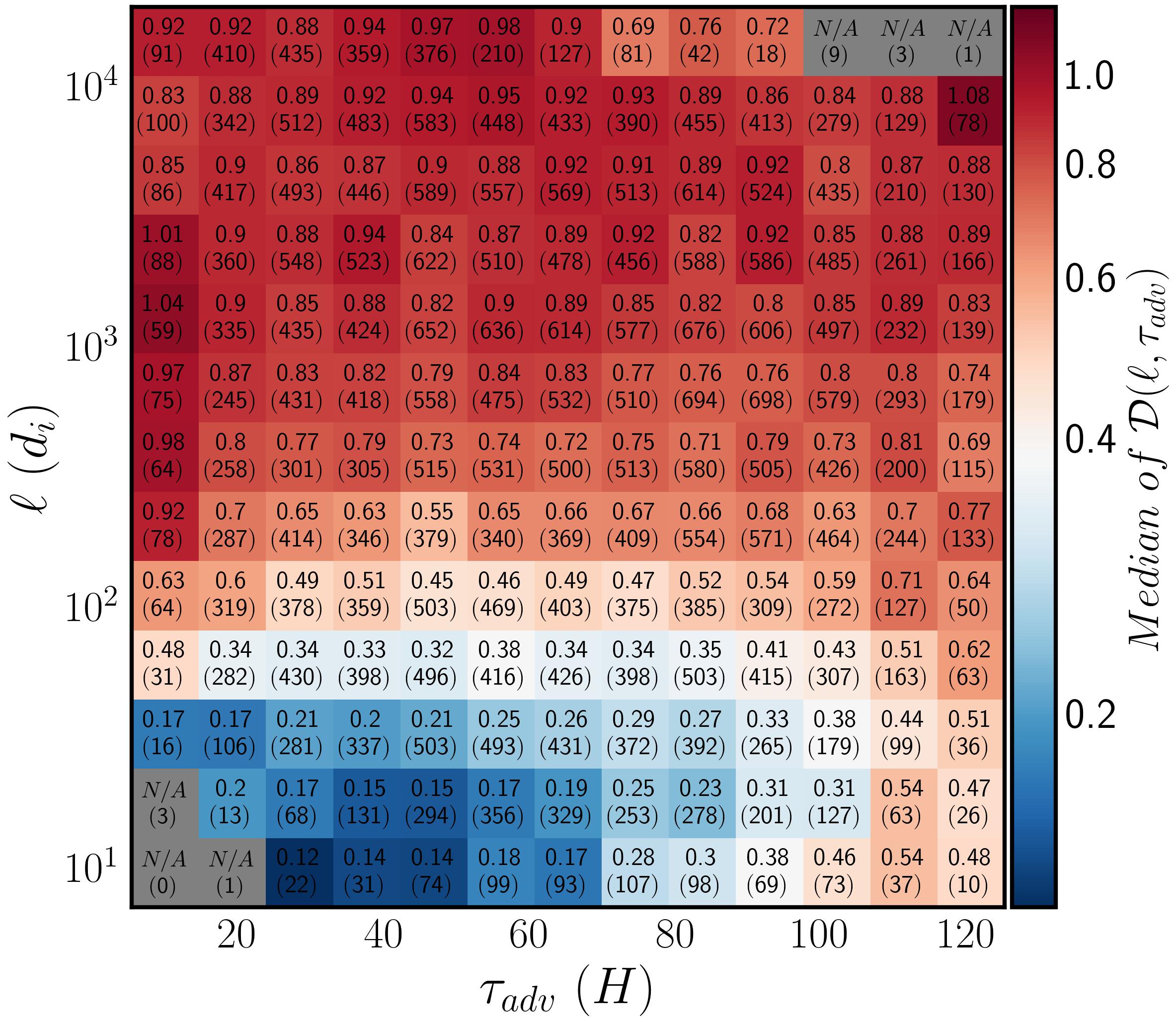}\label{fig:sfuncs_heatmap}

\caption{ To further quantify the divergence from the linear scaling predicted by the K41 theory of isotropic turbulence, $z_{k41}^{(q)} = q/3$, and as a measure of intermittency at a given scale, the quantity $\mathcal{D}(\ell, \ \tau_{adv}) = (\sum ( \frac{z(q, \  \ell)}{z(q=3, \ \ell)}-z_{k41}(q))^{2})^{1/2}$ is presented. The data points were binned according to $\ell$, and $\tau_{adv}$, and the median value inside each bin was calculated, which is reflected in the colors and written in the plot. The bracketed numbers in the plots are the number of data points inside each bin. Note that bins, including less than 10 data points, were discarded. }
\label{fig:SFuncs3}
\end{figure}


\subsubsection{Structure Functions (SFs).}\label{subsubsec:SF}

As outlined in Section \ref{subsec:Theory_SF}, a convenient way to describe the scaling variation of PDFs of the field fluctuations is by observing the deviation of the structure-function scaling exponents from the linear dependence on the order. Note that for a more direct comparison of the scaling exponents with the K41 linear scaling, we adapt the ESS hypothesis (see Section \ref{subsec:Theory_SF}). We thus proceed by dividing $\zeta(q)$ for the different lag ranges by $\zeta (3)$ for the respective range. Two $d =5$ hr-long intervals, sampled by PSP, and SolO, were randomly selected, and the structure functions $S^q_B(\ell)$ up to sixth order were calculated. Power-law fits have been applied on each $q_{th}$ order structure-function in the ranges between $(20-10^{2} d_i)$, $(10^{2}-10^{3} d_i)$ and $(10^{3}-10^{4} d_i)$ respectively. The resulting power-law exponents $\zeta(q)$ are  presented in Figure \ref{fig:SFuncs1}. The same process was then repeated for each of the 3100 intervals, and $\zeta(q)/\zeta(3)$ as a function of $q$ for the three different spatial scales is portrayed in Figure \ref{fig:SFuncs2}. It is important to note that the statistical accuracy of higher-order moments is affected by sample size. As a general rule, the highest order that can be computed reliably is $q_{max} = log(N) -1$, where N is the number of samples  \citep{Dudok_de_wit_Samples_Rule}. In this case, since the majority of the intervals contain  $N \sim 72000$ samples, we can estimate $q_{max} \approx 4$, meaning that higher-order statistics should be interpreted with caution. Consequently, a shaded gray area has been added to all the figures to indicate scaling exponents recovered for moments that were determined with questionable accuracy.


\begin{figure*}
\centering
\setlength\fboxsep{0pt}
\setlength\fboxrule{0.0pt}

\includegraphics[width=0.329\textwidth]{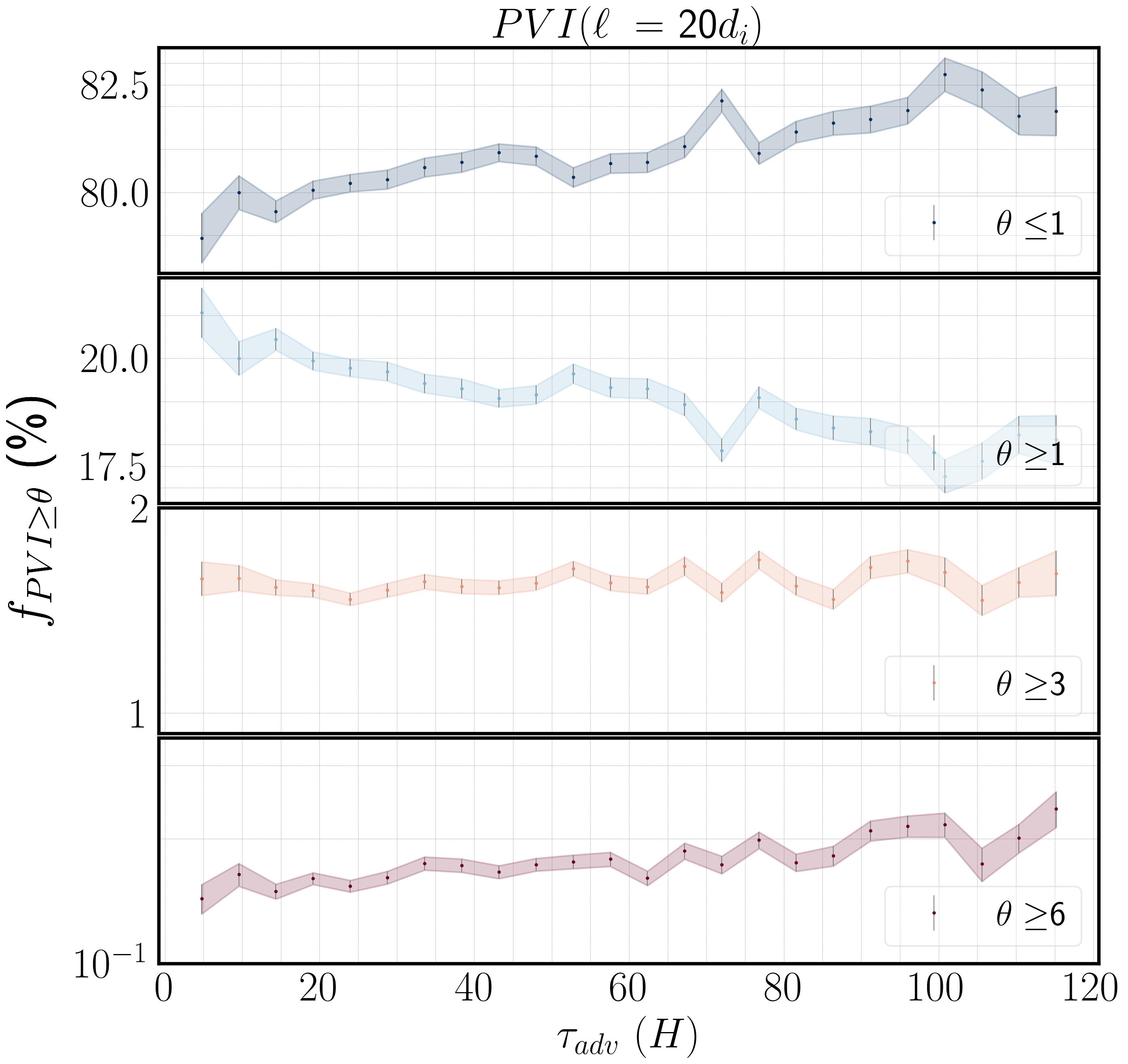}
\includegraphics[width=0.329\textwidth]{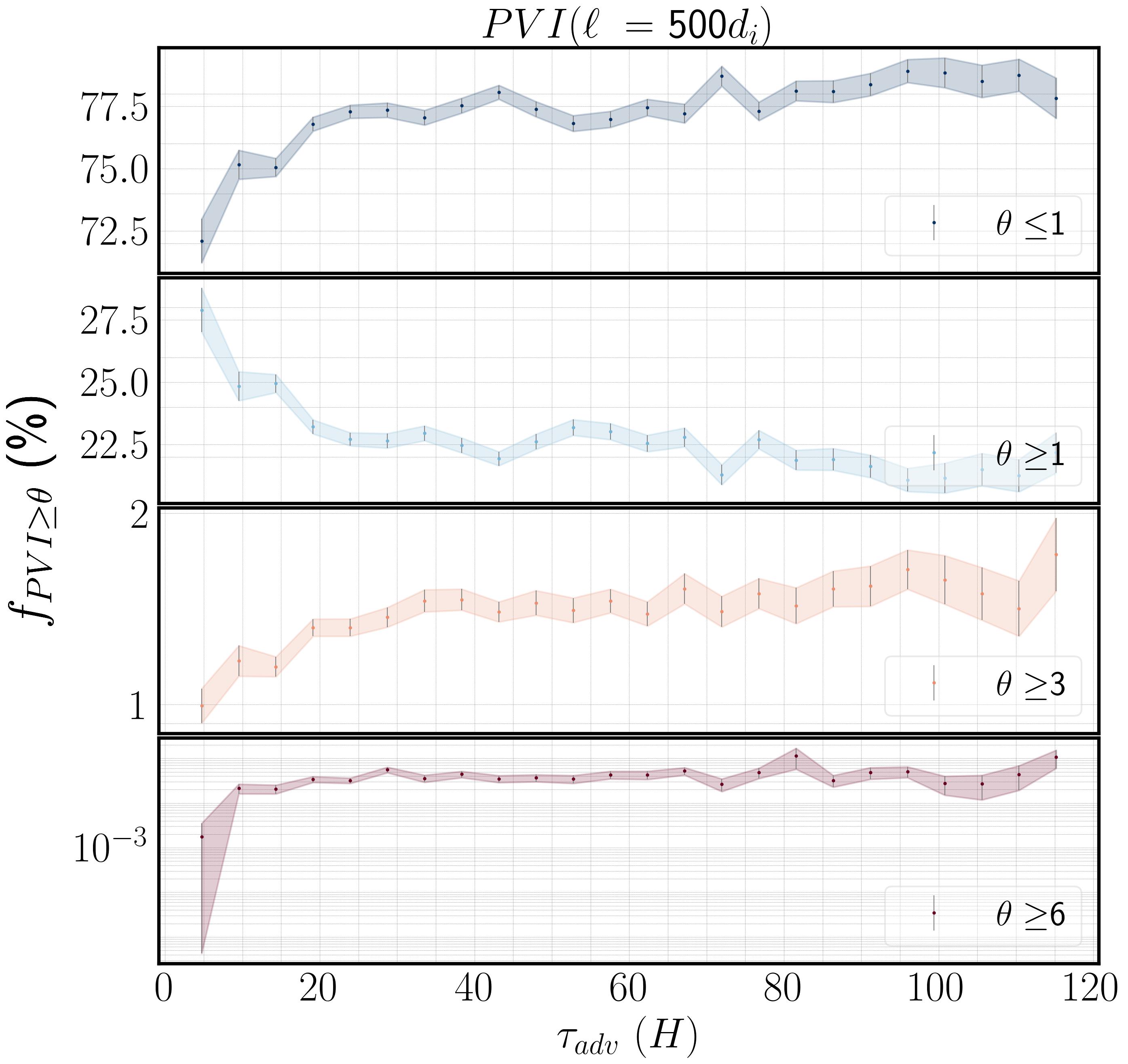}
\includegraphics[width=0.329\textwidth]{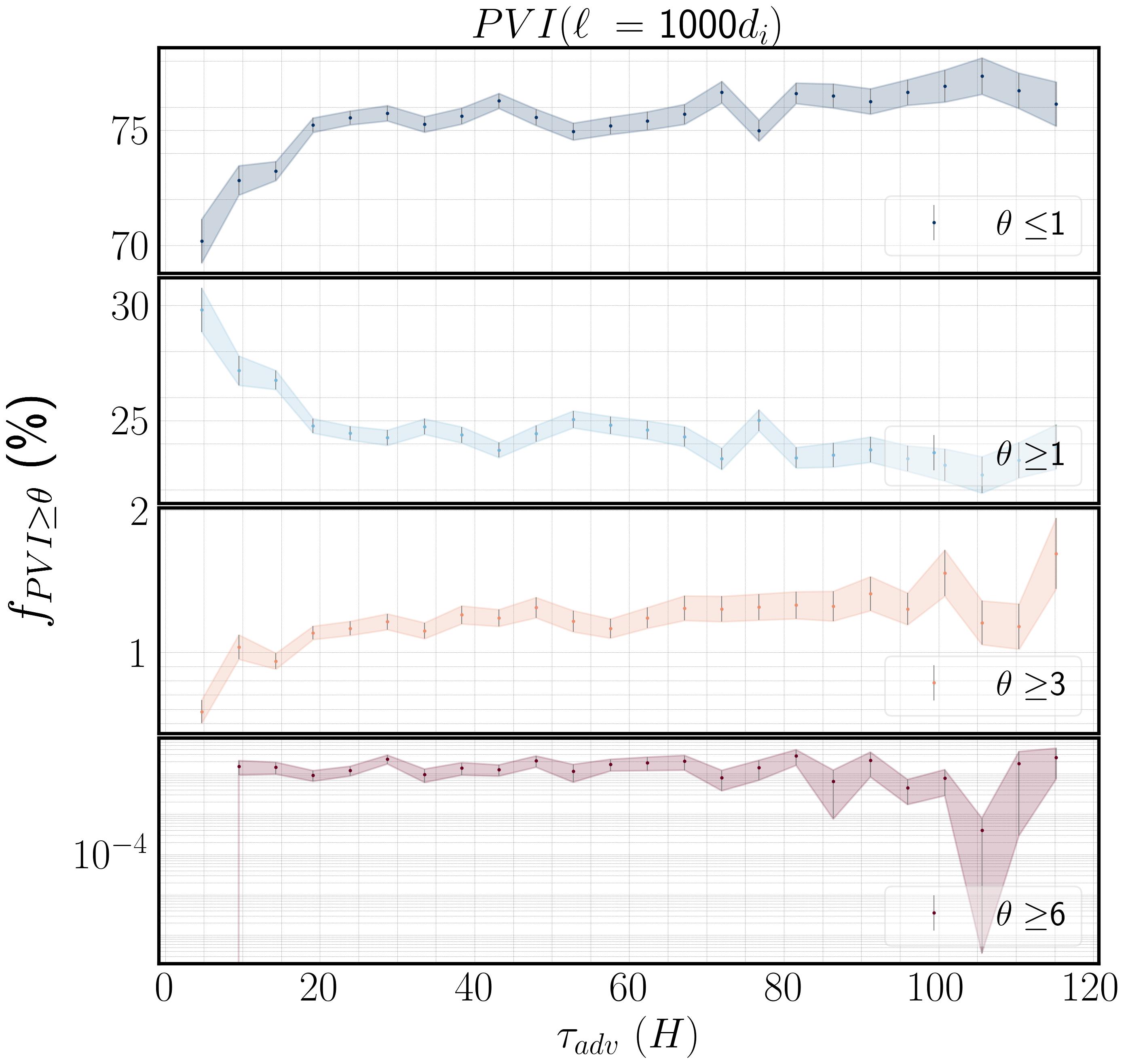}

\caption{Binned mean of the fraction of datapoints with $PVI$ value exceeding a given threshold ($f_{PVI\ge \theta}$, where $\theta = [1,3,6]$) shown as a function of advection time $\tau_{adv}$, different thresholds are shown in different colors.  Three spatial lags  normalized to the ion inertial length $d_i$ are considered, $\ell = 20, \ 500, \ 1000 \ d_{i} $.} 
\label{fig:PVI1}
\end{figure*}

In Figure \ref{fig:SFuncs2}a, the scaling exponents obtained by applying a best-fit linear gradient in log-log space over a window that spans between $(20 - 10^{2} d_i)$ are shown. Over this range of spatial scales, which, roughly speaking, corresponds to the transition region \citep{bowen_constraining_2020}, the scaling exponents obtain the highest observed values, indicating the presence of relatively stronger gradients in the magnetic field. Additionally, a roughly linear, i.e., monofractal, but super-Gaussian scaling at lower $\tau_{adv}$, which after normalization with $\zeta(3)$, closely resembles the K41 predicted curve, is obtained. However, as $\tau_{adv}$ increases, the lines exhibit a more concave behavior at large $q$. This result is in qualitative agreement with the results obtained through the SDK method and suggests the strengthening of transition range intermittency as a function of the advection time. In the inertial range, $(10^{2} - 10^{3} d_i)$ the normalized scaling exponents exhibit a concave scaling that strongly deviates from the K41 curve and does not show a dependence on the advection time of the solar wind. As a matter of fact, the concave scaling is observed at all times, indicating that the inertial range fluctuations exhibit a multifractal character even in the vicinity of the Sun. A similar behavior (i.e., multifractal scaling that does not evolve radially) is observed at yet larger,  nonetheless still inertial, scales between $(10^{3} - 10^{4} d_i)$. \par
To provide a quantitative context for the radial evolution of intermittency with respect to the advection time of the solar wind, the quantity $\mathcal{D}(\ell, \ \tau_{adv}) = (\sum ( \frac{z(q, \  \ell)}{z(q=3, \ \ell)}-z_{k41}(q))^{2})^{1/2}$ was estimated. This quantity is essentially the distance between two curves. The first curve corresponds to the scaling exponents $z(q,\  \ell)$ estimated for 5 hr-long intervals by applying a moving power-law fit to each of the structure functions up to order $q=6$, normalized by $z(q=3, \ \ell)$; the second curve corresponds to the K41 prediction for the scaling exponents $z_{k41}^{(q)})$. More specifically, the scaling exponents $\zeta(q, \ \ell)$ have been obtained by applying a moving power-law fit on each of the structure functions up to sixth order over a range of scales in the spatial domain $\ell \ \in \  [x_i, 3x_i]$, where $x_i$ is the starting point of the power-law fit; Each scaling exponent, $\zeta(q, \ell)$, has been normalized by $z(q=3, \ell)$ estimated over the same range of the respective interval. As a result, for a given interval and over a certain range of spatial scales, six normalized scaling exponents $\zeta(q, \ell)$/$z(q=3, \ell)$ have been obtained. Subsequently, the square of the deviation from the K41 prediction, $z_{k41}^{(q)})$, was estimated. Finally, the square root of the sum for $q = 1, ..., 6$ was calculated resulting in $\mathcal{D}(\ell, \tau_{adv})$. By sliding the moving fit window over the spatial domain, several estimates of $\mathcal{D}(\ell, \tau_{adv})$ have been obtained for different spatial scales over the same interval. The process was then repeated for all the available 5 hr-long intervals. Data points were subsequently binned according to $\ell$ and $\tau_{adv}$, and the median value of $\mathcal{D}(\ell, \  \tau_{adv})$ inside each bin was calculated, which is reflected in the colors. The results of this analysis are illustrated in Figure \ref{fig:SFuncs2}. The picture that emerges fits well with the evolution of SDK (see Figure \ref{fig:SDK1}) since the general trend is towards stronger intermittency at larger $\tau_{adv}$ for  $\ell \lesssim 100 d_i$ and no dependence on the advection time for $\ell \gtrsim 100 d_i$.
Obviously, the statistical trend is heavily influenced by the higher-order moments. The normalized scaling exponents, which relate to the moments of order $q \leq 3$, remain almost linear (see Figure \ref{fig:SFuncs2}), hence contributing a negligible amount to the difference between the two curves. It is also important to note that the same process was repeated by only taking into account scaling exponents up to $4^{th}$ order, and the results were found to be qualitatively similar.


\begin{figure}
\centering
\includegraphics[width=0.35\textwidth]{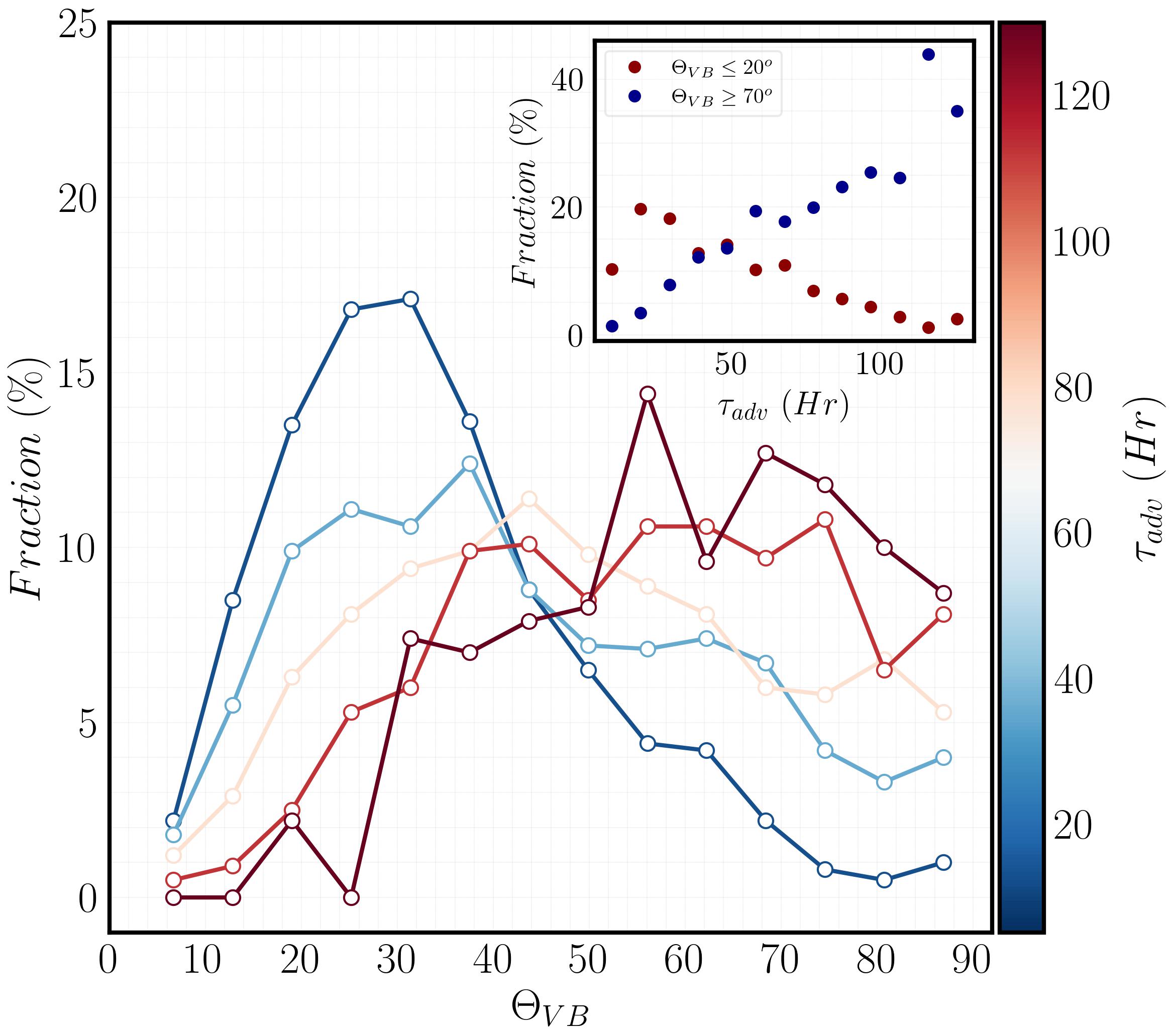}\label{fig:sfuncs_heatmap}

\caption{Five $\tau_{adv}$ bins are utilized to illustrate the fraction of $\Theta_{VB}$ angles for intervals that fall within each of the bins. Note that  the alignment angle is constrained to lie in the range between $0^{\circ}$
and $90^{\circ}$. For clarity, the inset figure shows the fraction of the dataset occupied by parallel ($\Theta_{VB} \leq\ 20^{\circ}$) and perpendicular intervals $\Theta_{VB} \geq \ 70^{\circ} $.}
\label{fig:theta_VB_histograms}
\end{figure}



\begin{figure*}
    \centering
    \setlength\fboxsep{0pt}
    \setlength\fboxrule{0.0pt}
    \includegraphics[width=0.329\textwidth]{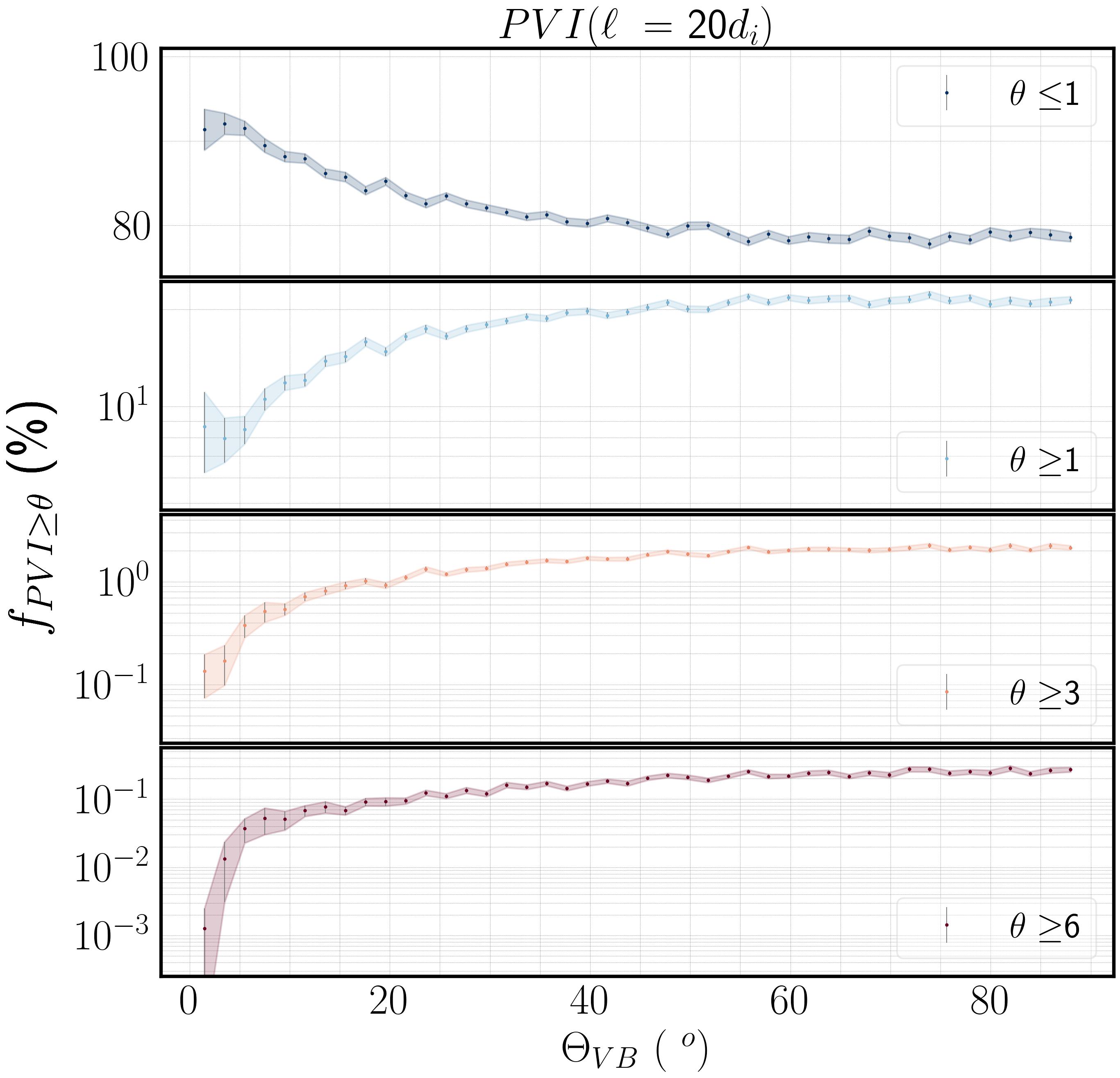}
    \includegraphics[width=0.329\textwidth]{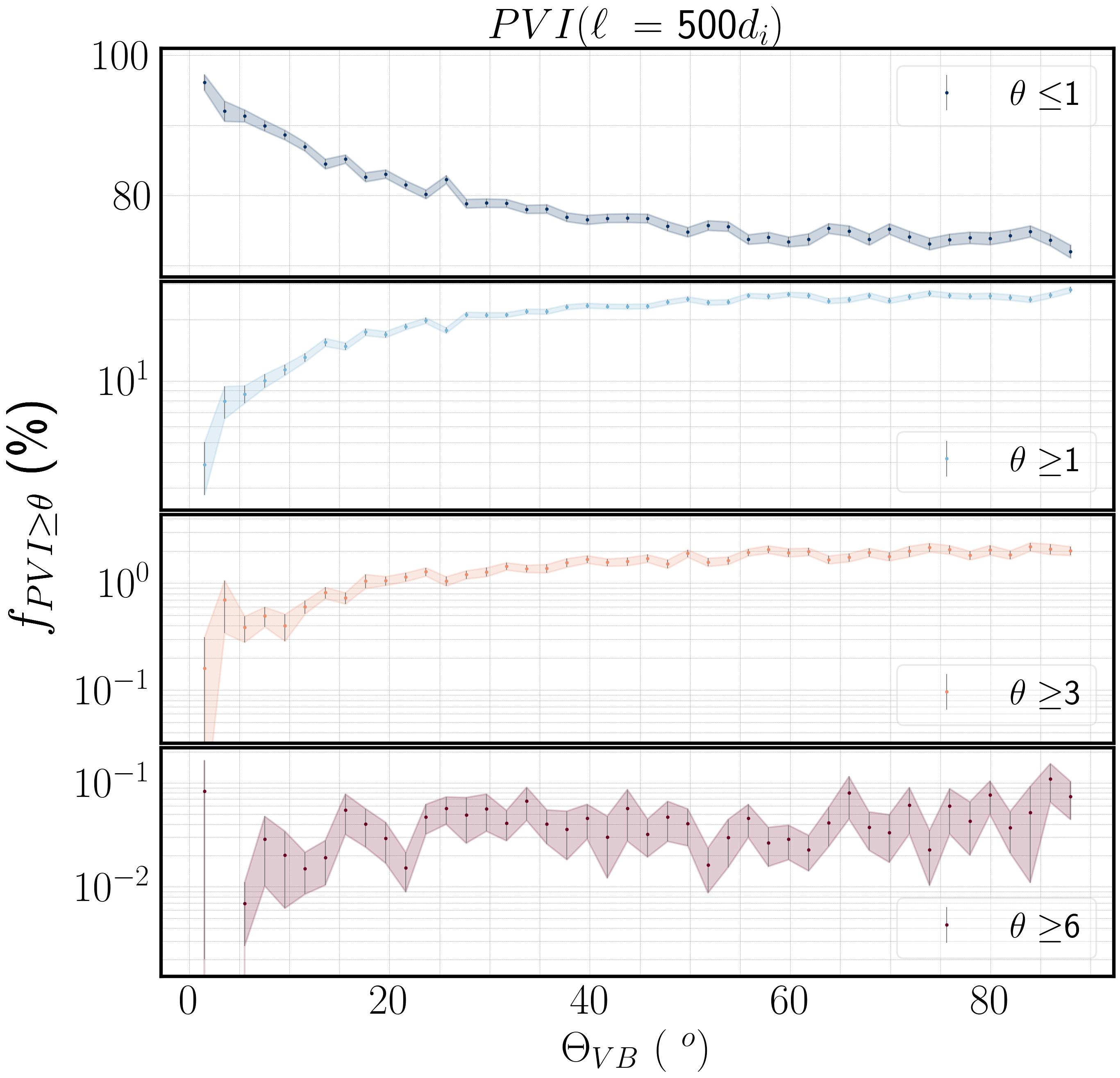}
    \includegraphics[width=0.329\textwidth]{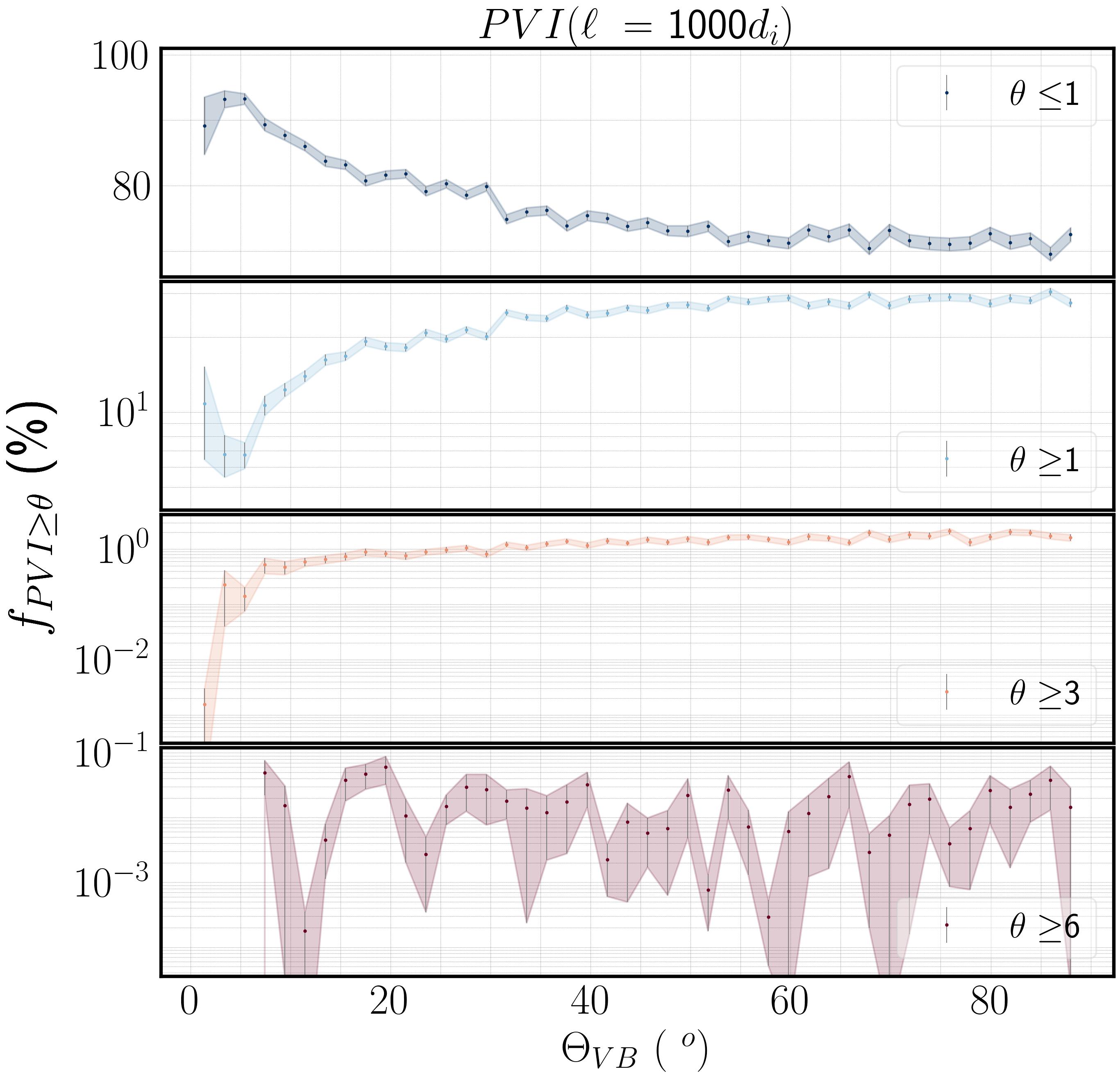}
    
    \caption{Binned mean of the fraction of PVI events $f_{PVI \ge\theta}\ (\%)$, where $\theta = 1, \ 3, \ 6$ as a function of the angle between solar wind and background magnetic field $\Theta_{VB}$. Three different lags are considered ($\ell = 1000, \ 500, \ 20 \ d_{i} $). A total number of $n =45$ bins have been used. } 
    
    \label{fig:PVI-theta-vb}
\end{figure*}

\subsubsection{Partial Variance of Increments (PVI).}\label{subsubsec:PVI}

In this section, we examine the radial dependence of intermittency by considering the evolution of the fractional volume with respect to the overall fluctuations occupied by coherent structures identified by means of the PVI method. As suggested in \citep{greco_intermittent_2008}, we consider a coherent structure any event for which the corresponding PVI index attains a value of $PVI > 2.5$. To estimate the PVI timeseries we follow the method outlined in Section \ref{subsubsec:PVI} and employ non-overlapping 10 hr-long intervals sampled by PSP and SolO throughout the inner heliosphere. To ensure estimating the PVI on a constant plasma scale, we adapt a lag normalized by the ion inertial length estimated at the respective interval (see Section. \ref{subsubsec:PVI}). After the PVI timeseries is estimated, data are further divided into 15-minute intervals, and the mean plasma advection time $\tau_{adv}$ for each interval is estimated. As a result of this process, a total number of  $\sim 30,000$, intervals of duration $d =15$ minutes are collected throughout the inner heliosphere. For a given interval, each data point is assigned a PVI value, and the fraction of the data points exceeding a certain PVI threshold $f_{PVI \geq \theta}$, where, $\theta= 1, \ 3, \ 6$, is estimated. In Figure \ref{fig:PVI1}a,b,c, we use 25 bins, to present the average value of $f_{PVI \geq \theta}$ per bin plotted against $\tau_{adv}$, for PVI estimated with a lag of $\ell$ = 20, 500, 1000 $d_{i}$, respectively. Uncertainty bars indicate the standard error of the sample. It is readily seen that as we move to smaller spatial lags (i.e., going from right to left panel), the fractional volume occupied by the extreme events is increasing. This result is consistent with the elevated probability density of extreme events when the PDFs of the magnetic increments are considered and indicates the presence of large gradients in the magnetic field at the smaller scales. In Figure \ref{fig:PVI1}a, for lag $\ell = 20 d_i$ and $PVI$ greater than 3 ($f_{PVI\ge 3}$, orange line), no clear trend is observed with $\tau_{adv}$. This result introduces a paradox to our analysis as it seems to contradict the evolution of $SDK$ at the smaller scales, shown in Figure \ref{fig:SDK1}a. However, when the evolution of $f_{PVI \ge 6}$ with lag $\ell=20 d_i$ (red line, panel a) is considered, a clear upward trend as a function of $\tau_{adv}$ is observed. Thus, a  natural hypothesis to explain the apparent contradiction is that the evolution of scale-dependent kurtosis (SDK) is dominated by the presence of the extreme events ($PVI \gtrsim 6$) that usually lay on the tails of the PDFs of normalized magnetic increments. The relationship between $f_{PVI \ge \theta}$ and $K_{max}$ is further examined in Appendix \ref{sec:Appendix}. From this analysis, we can understand that the fluctuations characterized by a smaller PVI index, albeit accounting for the bulk of the distribution of data points, have a relatively minor contribution to the final SDK values; On the other hand, even though the high PVI value events occupy only a small fraction of the dataset, they can significantly impact the behavior of SDK, due to the susceptibility of the fourth-order moment, found in the numerator of Equation \ref{eqn:SDK Definition}, to extreme increments. The trend of $f_{PVI\ge 6}$ is also consistent with the evolution of SDK at larger scales, shown in Figure \ref{fig:PVI1}b,c, as in both cases, the line practically remains flat as a function of $\tau_{adv}$. A different trend is observed for $f_{PVI \ge 3}$, and $\ell = 500, 1000 d_i$. In both cases, an abrupt increase in $f_{PVI \ge 3}$ (yellow line) is observed up to $\sim 20-25$ hours. This feature is of particular importance, as it could be related to the crossing of PSP through the Alfv\'en region and is further discussed in Section \ref{sec:Summary}. Beyond this point, a slight upward trend is observed at both lags for subsequent times. 
 Another interesting feature in Figures \ref{fig:PVI1}a,b,c is the evolution of fluctuations with $PVI < 1$, as in all three cases a monotonic increase of $f_{PVI \leq 1}$ with $\tau_{adv}$ is observed.  Since both $f_{PVI \leq 1}$ and $f_{PVI \leq 3}$ are following an upward trend, we can understand that the fluctuations in the $1 \lesssim PVI \lesssim 3$ are gradually getting depleted as the solar wind expands. In particular, for $\ell$, the depletion process is gradual with a decrease of $\approx 2.5 \ \%$, observed between 5 and 130 hours. On the other hand, at the largest scales following an abrupt decrease of $\approx 5 \ \%$  up to $\approx 25 $H, the $f_{PVI<1}$ practically remains constant over the ranges examined.





\begin{figure*}
    \centering
    \setlength\fboxsep{0pt}
    \setlength\fboxrule{0.0pt}

    \includegraphics[width=0.443\textwidth]{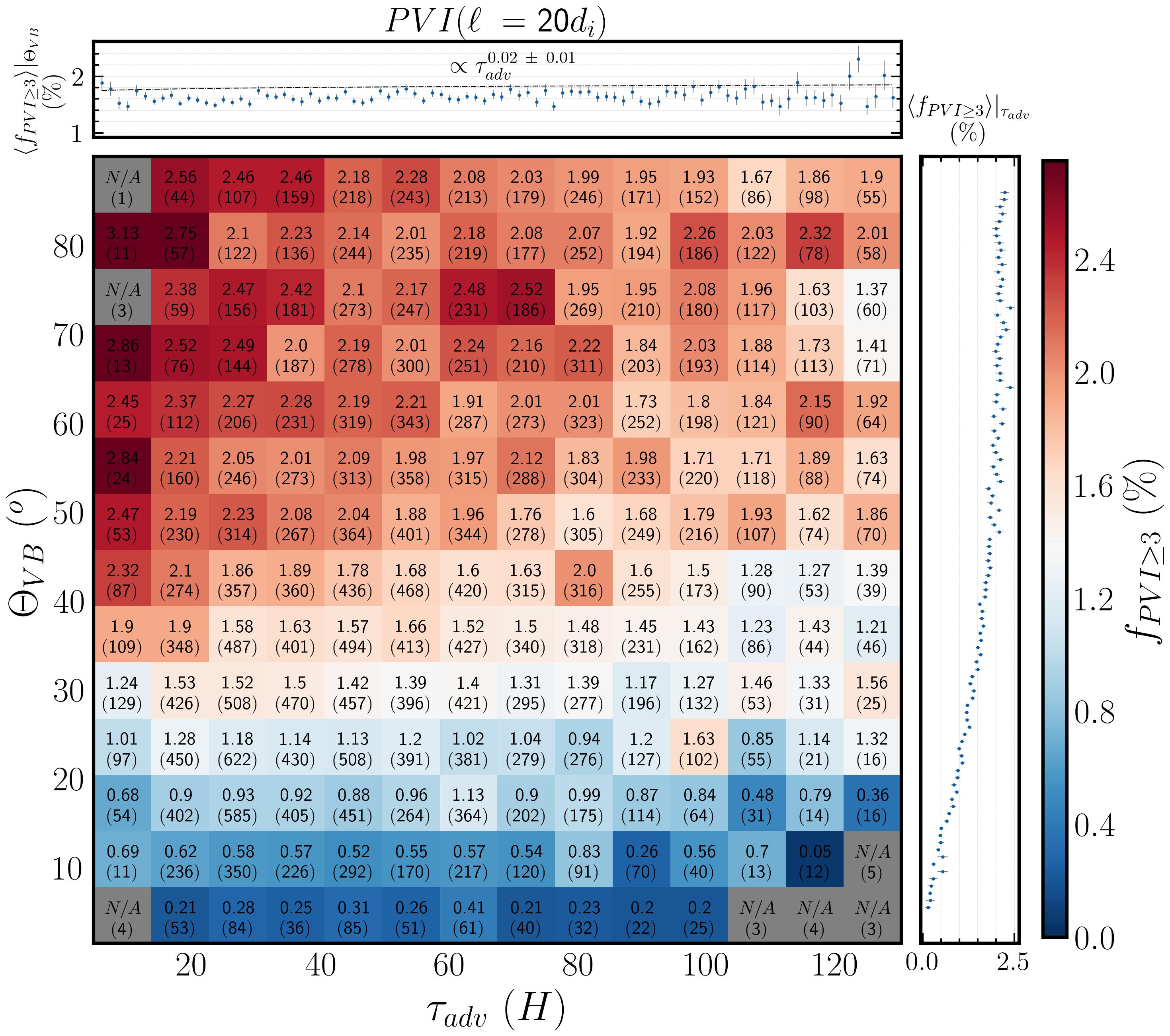}
    \includegraphics[width=0.45\textwidth]{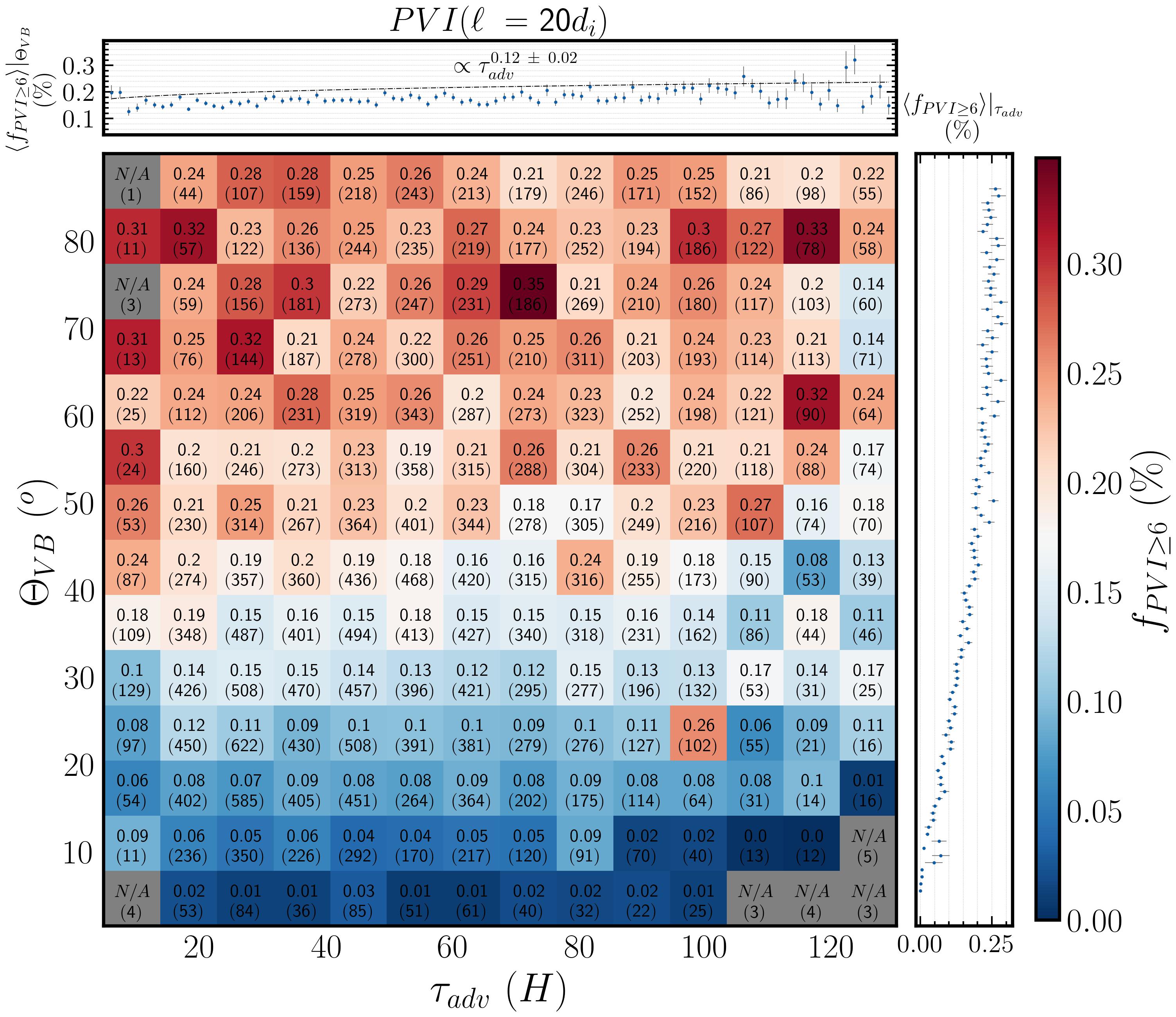}

    \caption{ Binned mean of the fraction of datapoints with $PVI$ value exceeding a given threshold, $f_{PVI\ge \theta}$, where $\theta = 3$ (left panel) and $\theta = 6$ (right panel) shown as a function of advection time $\tau_{adv}$ and $\Theta_{VB}$ angle.  The PVI timeseries was estimated with a lag, $\ell = 20 \ d_{i} $. The top subplot shows the averaged fraction of coherent structures, over $\Theta_{VB}$, $\langle f_{PVI \geq \theta} \rangle|_{\Theta_{VB}}$. Power-law fits are also shown as black-dashed lines. Similarly, the right subplot shows the averaged fraction over $\tau_{adv}$, $\langle f_{PVI \geq \theta} \rangle|_{\tau_{adv}}$. }
    \label{fig:PVI-tadv-theta-vb-heatmap}
\end{figure*}

\subsubsection{Angle between Solar Wind Flow and Magnetic Field $\Theta_{VB}$}\label{subsec:thetavb}

Two important factors need to be considered when studying the radial evolution of intermittency in the solar wind: (a) MHD turbulence in the solar wind has a well-known tendency to develop and sustain several manifestations of anisotropy, e.g., wavevector anisotropy, variance anisotropy, etc. \citep{2015RSPTA.37340152O}. One of these is the anisotropy in magnetic field intermittency introduced by the presence of the background solar wind flow. For parallel intervals (i.e., $\Theta_{VB} \approx 0^{\circ}$, or equivalently $\Theta_{VB} \approx 180^{\circ}$), the statistical signature of the magnetic field fluctuations is that of a non-Gaussian globally scale-invariant process, in contrast to multi-exponent statistics observed when the local magnetic field is perpendicular to the flow direction  \citep{horbury_anisotropic_2008, osman_intermittency_2012}. (b) Because of the conservation of magnetic flux (Parker Spiral), the radial component of the magnetic field decreases faster than the transverse component.

\begin{figure}

\includegraphics[width=0.44\textwidth]{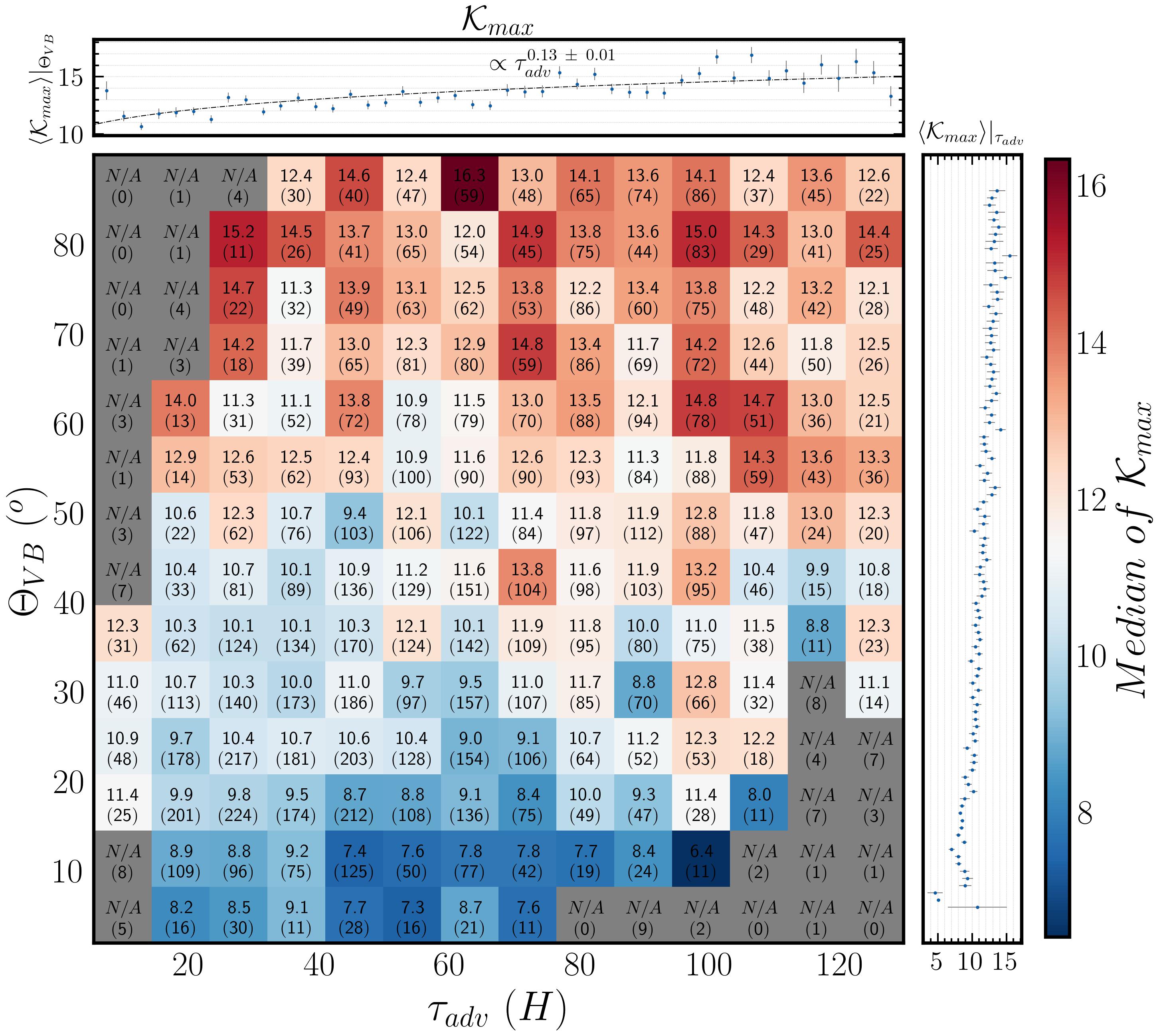}
\caption{Scale-dependent kurtosis of the magnetic field as a function of $\Theta_{VB}$ and  $\tau_{adv}$. The numbers indicate the median value of kurtosis within each bin. The top subplot shows $K_{max}$, over $\Theta_{VB}$, $\langle K_{max} \rangle|_{\Theta_{VB}}$. Power-law fits are also shown as black-dashed lines. Similarly, the right subplot shows the averaged fraction over $\tau_{adv}$, $\langle K_{max} \rangle|_{\tau_{adv}}$. }
\label{fig:SDK_thetaVB_1}
\end{figure}



 Consequently, as radial distance increases, so does the number of perpendicular intervals. On the other hand, as shown in the inset of Figure \ref{fig:theta_VB_histograms}, the fraction of parallel/anti-parallel intervals is monotonically decreasing. Therefore, for a complete understanding of the radial evolution of intermittency in the solar wind, an analysis that takes into account both $\tau_{adv}$ and $\Theta_{VB}$ is required. In this section, we examine the radial evolution of anisotropic intermittency by means of the PVI and SDK methods. As a first step, the PVI method and the 15-min intervals described in Section \ref{subsec:Results_PVI} are adapted. Having confirmed that the anisotropy is symmetric with respect to $\Theta_{VB}=90^{\circ}$, we proceeded by not distinguishing between parallel and anti-parallel directions. As a result, intervals with an estimated $\Theta_{VB}^{init} \geq 90^{\circ}$ have been recast to $\Theta_{VB} = 180^{\circ} -\Theta_{VB}^{init}$. We, therefore, require that the alignment angles lie within a range between $0^{\circ}$ and $90^{\circ}$ degrees. In  Figures \ref{fig:PVI-theta-vb}a,b,c, the fraction of PVI events at a given PVI threshold, $f_{PVI \geq \theta}$, is plotted against the $\Theta_{VB}$ angle for PVI estimated with lag $\ell = 20, \ 500, \ 1000 \ d_{i} $, respectively. For clarity, data have been binned into 45 linearly spaced bins in the $\Theta_{VB}$ domain, and each dot indicates the mean value of $f_{PVI\geq \theta}$ within the bin. Error bars are also shown, indicating the standard error of the mean. For $\ell=20d_i$ (Figure \ref{fig:PVI-theta-vb}a), the fraction of random fluctuations with $PVI < 1$ is rapidly decreasing as intervals with greater $\Theta_{VB}$ angles are considered. The opposite trend is observed in the fraction of magnetic increments with $PVI \geq 1$. As a matter of fact, the anisotropy grows stronger as higher PVI thresholds are considered. For instance, an increase in $f_{PVI \ge \theta}$ of at least one and two orders of magnitude is recovered between the lowest and highest $\Theta_{VB}$ angles for PVI thresholds $\theta = 3, \ 6$, respectively. Note, however,  that regardless of the PVI threshold value, the increasing trend is halted at $\Theta_{VB} \approx 50^{\circ}$. Beyond this point, no statistically significant differences in $f_{PVI\geq \theta}$ are observed at the largest $\Theta_{VB}$ angles. A similar degree of anisotropy, as a function of $\Theta_{VB}$, is recovered for larger lags ($\ell=500, 1000 d_i$) shown in Figure \ref{fig:PVI-theta-vb}b,c, respectively. Note, however, a  deviation from the trend for $f_{PVI \geq 6}$ indicating that the degree of anisotropy is lessened at progressively larger spatial scales.
 
In Figure \ref{fig:PVI-tadv-theta-vb-heatmap}, the evolution of anisotropic intermittency is examined as a function of $\tau_{adv}$. For this reason, the data points were binned according to $\Theta_{VB}$ and $\tau_{adv}$, and the mean value inside each bin was calculated, which is reflected in the colors and presented in the plot. The bracketed numbers in the plots are the number of data points inside each bin. Note that bins, including less than 10 data points, were discarded. In Figure \ref{fig:PVI-tadv-theta-vb-heatmap}a, the dependence of $f_{PVI \geq 3}$ as a function of $\Theta_{VB}$ and $\tau_{adv}$ for lag $\ell = 20 d_i$ is illustrated. 
 Two major but contradicting conclusions can be drawn from this figure: 
 (1) When the evolution of intervals that belong to the same $\Theta_{VB}$ bin is considered, a monotonic decrease in $f_{PVI \ge 3}$ is observed for all $\Theta_{VB}$ rows. (2) The average $f_{PVI \ge 3}$, with regard to $\Theta_{VB}$, shows a negligible, slightly positive trend as a function of $\tau_{adv}$. This seemingly inconsistent result can be addressed by considering the radial evolution of the Parker spiral. Closer to the Sun, the solar wind speed and background magnetic field tend to be aligned, i.e., the intervals tend to concentrate around $\Theta_{VB}\sim 0^\circ(180^\circ)$. As we move further away from the Sun, this angle shifts towards the perpendicular direction, i.e., $\Theta_{VB}\sim 90^\circ$. However, as shown in Figure \ref{fig:PVI-tadv-theta-vb-heatmap}, perpendicular intervals are typically associated with higher $f_{PVI \geq 3}$ values. Thus, despite the gradual decrease in the fraction of coherent structures with $PVI \ge 3$ as a function of $\tau_{adv}$ for a constant $\Theta_{VB}$ angle, on average, the fraction of the entire dataset shows signs of a  very subtle increase. When considering the evolution of coherent structures of $PVI, \ge 6$, Figure \ref{fig:PVI-tadv-theta-vb-heatmap}b, a slightly different evolution may be noticed. In particular, not a clear trend is observed for intervals of constant $\Theta_{VB}$. Additionally, as pointed out in Figure \ref{fig:PVI-theta-vb}a the degree of anisotropy with regards to the $\Theta_{VB}$ angle is strengthened when higher PVI thresholds are considered. Therefore, by applying the same logic as outlined before for $PVI \ge 3$, we can explain the apparent increase in the fraction of coherent structures as a function of $\tau_{adv}$.
 
\begin{figure}

\includegraphics[width=0.45\textwidth]{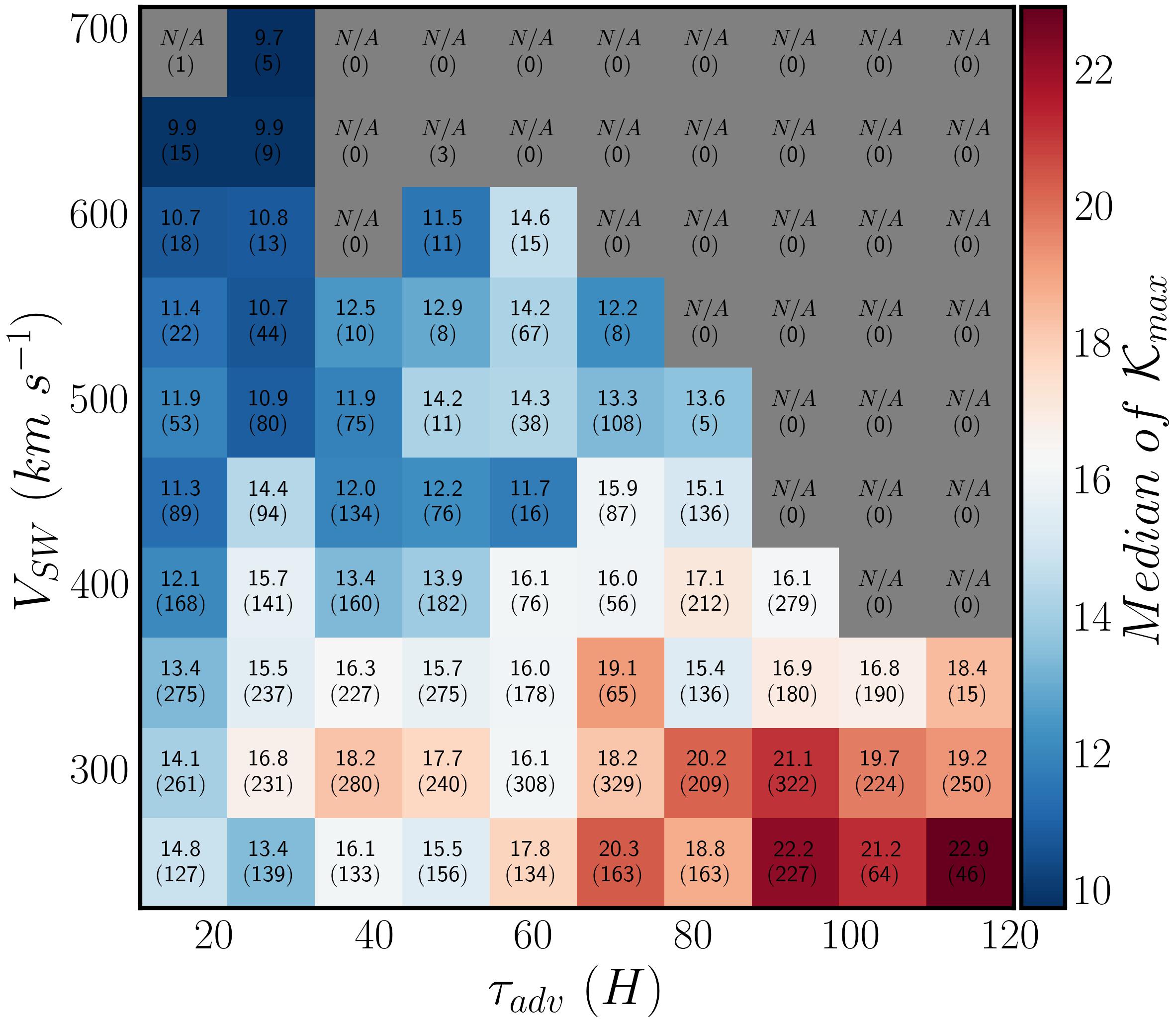}
\caption{(a) Scale-dependent kurtosis of the magnetic field as a function of solar wind speed $V_{SW}$ and SW advection time $\tau_{adv}$. The numbers indicate the median value of kurtosis within each bin. }
\label{fig:VSW_intermit1}
\end{figure}

 We move on to examine the evolution of $K_{max}$ as a function of $\tau_{adv} $ and $\Theta_{VB}$. In order to mitigate the effects of mixing different types of solar wind, the duration of the intervals used has been reduced to $d = 30 $ minutes. It is important to note that even though the radial trend of kurtosis is not affected  (i.e., the maximum of the kurtosis is observed to increase with increasing $\tau_{adv}$ regardless of interval size), the curves are shifted vertically to larger values when larger averaging windows are considered. This may be attributed to the fact that by increasing the interval size, more and more extreme events are taken into consideration during the averaging process. Since these events have been shown to strongly affect the SDK (see Section \ref{subsec:Results_PVI}), an increase of SDK for larger averaging windows is to be expected. The results of this analysis are illustrated in Figure \ref{fig:SDK_thetaVB_1}. As expected, $K_{max}$ follows a qualitatively similar trend with $f_{PVI \ge 6}$. More specifically, intervals for which the magnetic field and solar wind speed tend to be aligned exhibit lower $K_{max}$ values. Moreover, no clear trend may be observed with $\tau_{adv}$ when examining intervals with a similar $\Theta_{VB}$.

 \par
 As a result of our analysis, it is apparent that understanding the physical mechanisms driving the evolution of intermittent properties in the magnetic field of the solar wind requires making a distinction between the effects of mixing strongly and less intermittently perpendicular and parallel intervals, respectively, as opposed to the evolution of turbulence during the expansion due to the local plasma dynamics.

\subsection{Dependence of intermittency in plasma parameters.}\label{subsec:intermittency_plasma_params}

\begin{figure}
\includegraphics[width=0.45\textwidth]{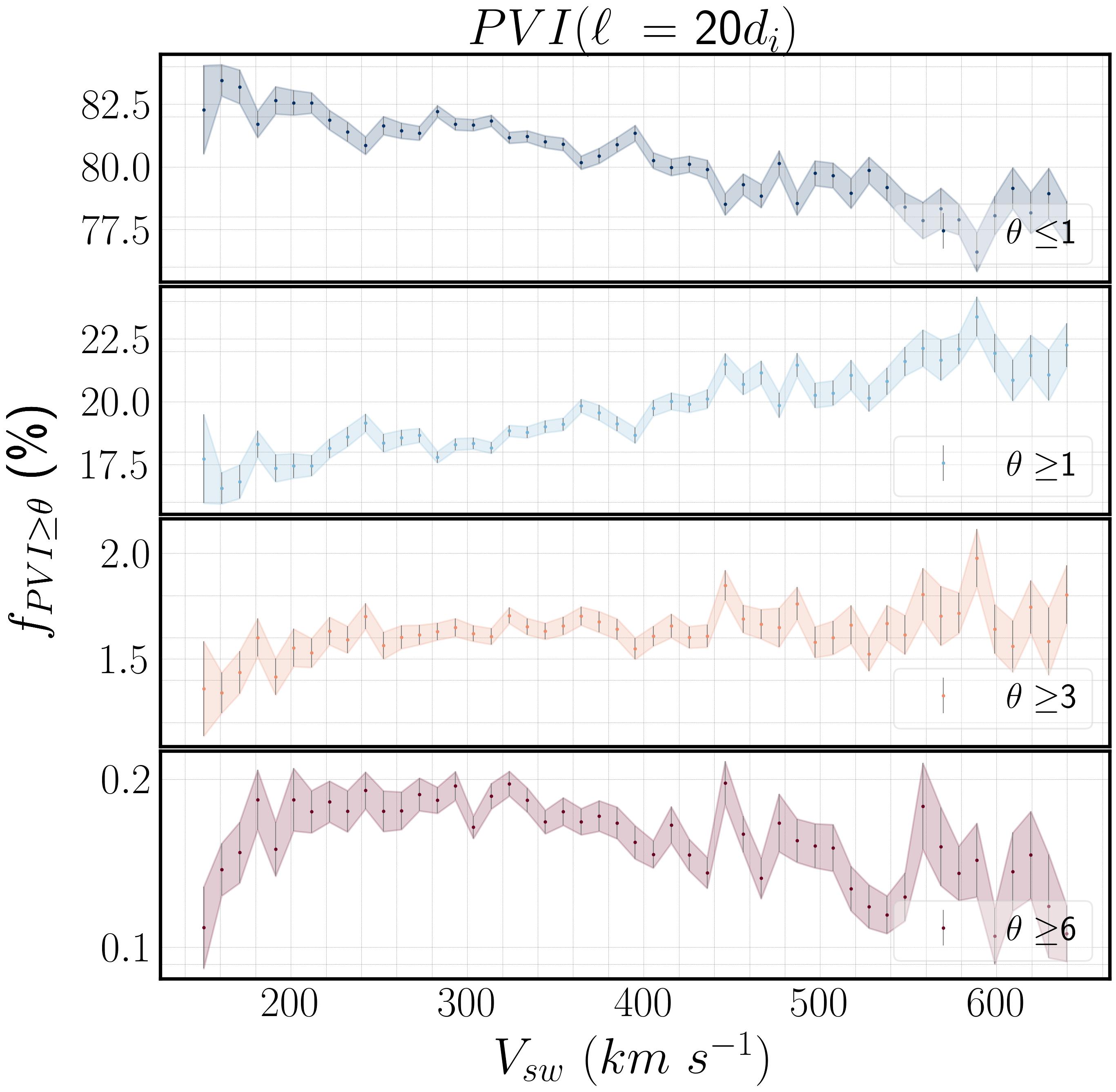}

\caption{Fraction of datapoints with $PVI$ value exceeding a given threshold, $f_{PVI\ge \theta}$, where $\theta  =1, \ 3, 6$ as a function of solar wind speed, $V_{SW}$.}
\label{fig:VSW_intermit2}
\end{figure}

\begin{figure*}
    \centering
    \setlength\fboxsep{0pt}
    \setlength\fboxrule{0.0pt}
    
    \includegraphics[width=0.45\textwidth]{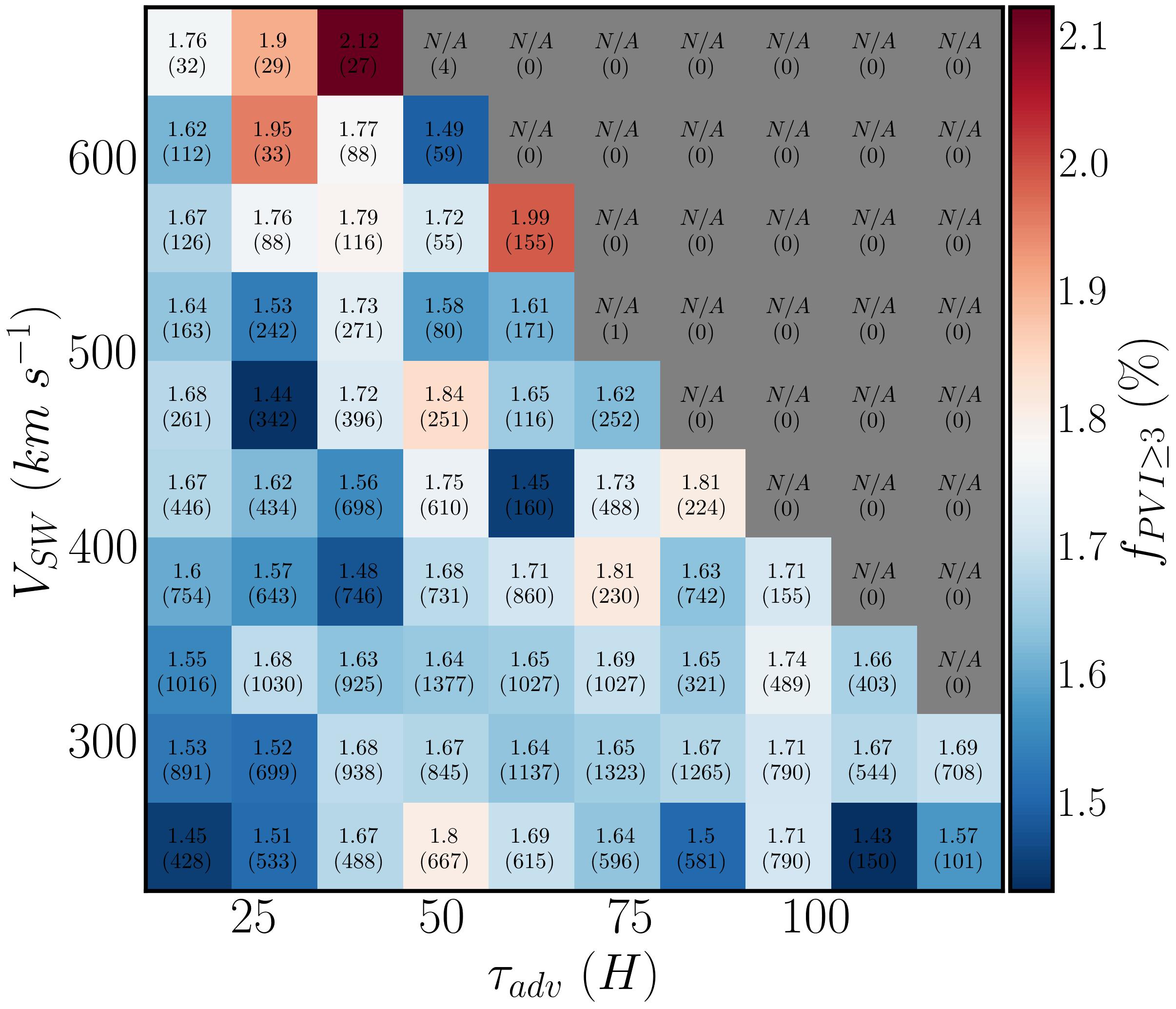}
    \includegraphics[width=0.45\textwidth]{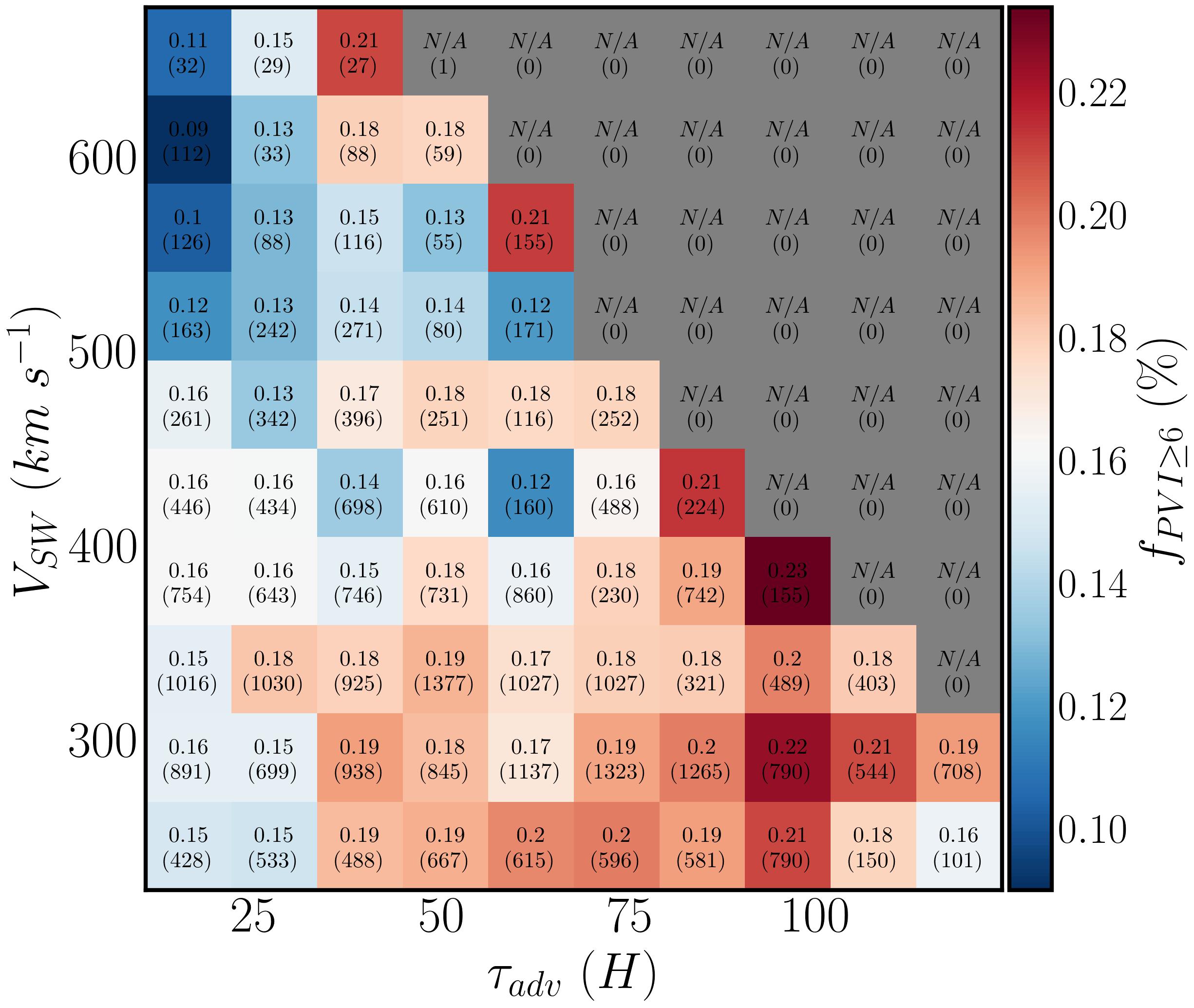}

\caption{Fraction of datapoints with $PVI$ value exceeding a given threshold, $f_{PVI\ge \theta}$, where (a) $\theta  =3$, (b) $\theta  =6$, as a function of $V_{SW}$ and $\tau_{adv}$. The numbers indicate the median value of $f_{PVI\ge \theta}$ and the bracketed numbers show the number of events within each bin.}
\label{fig:VSW_intermit3}
\end{figure*}

\subsubsection{Solar Wind Speed}\label{subsubsec:solar wind speed}

 In this section, the relationship between solar wind speed $V_{SW}$ and magnetic field intermittency is investigated. Similarly to Section \ref{subsec:thetavb}, to mitigate the effects of mixing different types of solar wind, the duration of the intervals used has been reduced to $d = 30 $ minutes. Also, note that the intervals that are associated with fast solar wind $V_{SW} \gtrsim 600 \ km \ s^{-1}$ comprise only a minor fraction of our dataset. Moreover, the majority of these intervals were observed during the latest perihelia of PSP and thus in proximity to the Sun. Taking these arguments into account, we can understand that the study of the radial evolution of the fast wind is not feasible with our current dataset. Nevertheless, the study of intermittency properties as a function of $V_{SW}$ is still possible since a considerable number of intervals with $V_{SW}$ in the range  $200 \  km \ s^{-1} \lesssim V_{SW} \lesssim \  600 km \ s^{-1}$ have been sampled by both PSP and SolO throughout the inner heliosphere. We begin our analysis by considering the relationship between $K_{max}$ with $\tau_{adv}$ and $V_{SW}$, estimated for respective intervals. The results of this analysis are presented in Figure \ref{fig:VSW_intermit1}. 
In accordance with \citep{bruno_radial_2003, Weygand_kurtosis}, we find that the kurtosis for the magnetic fluctuations in the slow solar wind exhibit higher peaks when compared to those of the fast solar wind. As a matter of fact, $K_{max}$ almost monotonically decreases as a function of $V_{SW}$ when intervals sampled at the same $\tau_{adv}$ column are considered. Additionally, regardless of the solar wind speed, $K_{max}, $ increases as a function of $\tau_{adv}$. This result comes in disagreement with \citep{bruno_radial_2003}, as it indicates the radial strengthening of intermittency regardless of the solar wind speed. \par

Additionally, the relationship between the fractional volume occupied by coherent structures, $f_{PVI \geq \theta}$, identified by using the PVI method (see Section \ref{subsec:Results_PVI}), and $V_{SW}$ is examined. The results of this analysis are illustrated in Figure \ref{fig:VSW_intermit2}, for PVI threshold $\theta =1, \ 3,\ 6$. It is readily seen that fast solar wind is characterized by an elevated number density of magnetic increments with PVI index greater than unity, $\theta \geq 1$ (cyan line). Strictly speaking, and following the definition of \citep{greco_intermittent_2008}, only events of $PVI \gtrsim 2.5$ correspond to coherent structures and consequently strengthen the intermittent character of the magnetic fluctuations. However, this result is of interest as it was recently shown \citep{sioulas_statistical_2022} that the number density of structures with PVI index greater than unity $f_{PVI \geq 1}$ is very strongly correlated with the temperature of protons $T_{p}$ in the solar wind. At the same time, one of the clearest correlations between plasma parameters in the solar wind is the one between proton temperature with solar wind speed  \citep{Perrone19:radial}. It is thus quite probable that events of PVI index greater than unity not only contribute to magnetic energy dissipation and the heating of the ambient plasma environment but at the same time are partly responsible for the acceleration of the solar wind. Moving on and considering the events of $\theta \geq 3$ (shown in yellow), a picture that contradicts our conclusions from the SDK analysis outlined earlier emerges. In particular, within the error bars, no statistically significant differences in the fractional volume occupied by coherent structures can be observed between fast and slow solar wind streams. As a matter of fact, one could even argue that a slight increase of $f_{PVI \geq 3}$ can be observed with increasing solar wind speed. A different picture emerges when we consider the highest PVI threshold $PVI \geq 6$. More specifically, $f_{PVI \geq 6}$ is progressively reduced when faster solar wind streams are considered. This result provides a natural explanation for the lower SDK peaks observed at faster solar wind streams since, as already discussed in Section \ref{subsec:Results_PVI}, the number density of PVI greater than 6 events, $f_{PVI \geq 6}$ is tightly correlated with $K_{max}$. We move on to investigate the evolution of $f_{PVI \geq \theta}$ as a function of $V_{SW}$ and $\tau_{adv}$. The results are presented in Figure\ref{fig:VSW_intermit3}  for $PVI \geq 3 $, and $PVI \geq 6$ respectively. For $PVI \geq 3$, though on average slightly higher values of $f_{fVI \geq 3}$, may be observed at greater  $\tau_{adv}$, there is, strictly speaking, not a clear horizontal trend. On the contrary, for $PVI \geq 6$, an increasing trend is observed for most of the rows (i.e., streams of similar solar wind speed) in qualitative agreement with the increasing $K_{max}$ reported in Figure \ref{fig:VSW_intermit2}.

\begin{figure*}
\centering
\setlength\fboxsep{0pt}
\setlength\fboxrule{0.0pt}
\includegraphics[width=0.45\textwidth]{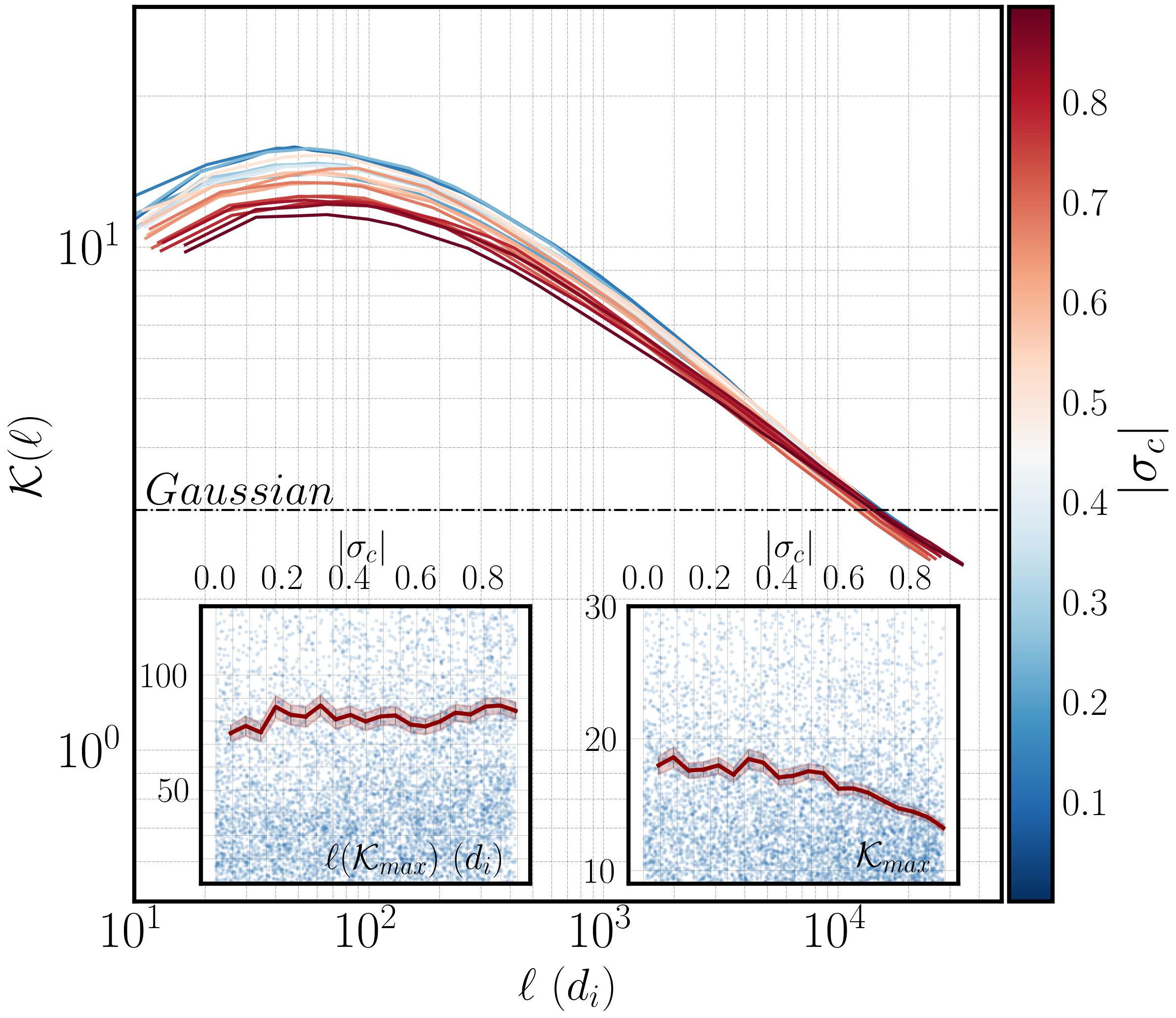}
\includegraphics[width=0.445\textwidth]{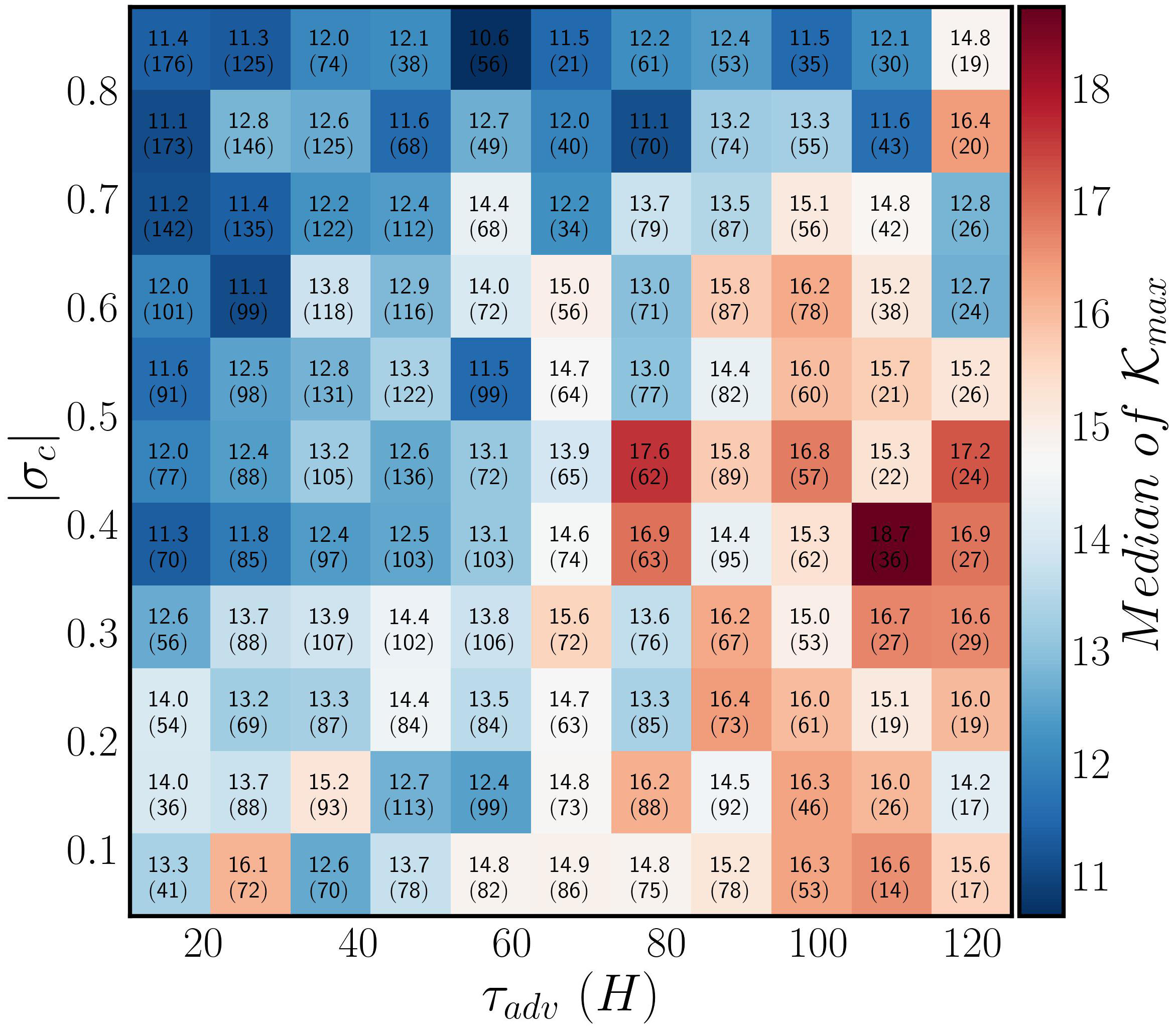}

\caption{(a) Scale-dependent kurtosis of the magnetic field as a function of scale in units of ion inertial length $d_{i}$ and SW advection time $\tau_{adv}$. The numbers indicate the mean value of kurtosis within each bin. (b) Evolution of SDK with $\tau_{adv}$. Each line represents the average of 100 intervals that fall within the same $\tau_{adv}$ bin.
The inset scatter plots show (i) the scale, in units of $d_i$,  at which the kurtosis attains the maximum value  (ii) The maximum kurtosis value estimated for the individual 5H intervals. The binned mean of the two quantities is shown (red line).}
\label{fig:sigma_c}
\end{figure*}

\subsubsection{Normalized cross helicity.}\label{subsubsec:sigma_c}

In this section, the correlation between the normalized cross-helicity and intermittency, as indicated by the scale-dependent kurtosis (SDK) of the magnetic field magnitude, is examined. The normalized cross-helicity $\sigma_c$ is defined as:
\begin{eqnarray}
    \sigma_c = \frac{\langle \boldsymbol{z}_+^2\rangle-\langle \boldsymbol{z}_-^2\rangle}{\langle \boldsymbol{z}_+^2\rangle+\langle \boldsymbol{z}_-^2\rangle}
\end{eqnarray}
where $\boldsymbol{z}_\pm=\boldsymbol{v} \pm \boldsymbol{v}_{a}$, $\boldsymbol{v}_{a}=\delta \boldsymbol{B}/\sqrt{\mu_0 m_p \langle n_p\rangle}$, $\delta \boldsymbol{B}=\boldsymbol{B}-\langle B \rangle$, and $\langle\ \rangle$ is the ensemble average. Note that for this analysis, the length of the interval has been reduced to $d=30min$, to ensure that $\sigma_c$ does not vary significantly within the interval. For each interval, the median of $\sigma_c$ has been estimated, and intervals with standard deviation of $\sigma_c$ greater than 0.2 have been discarded. In Figure \ref{fig:sigma_c}a, the dependence of SDK as a function of $|\sigma_c|$ is illustrated. Note that each line corresponds to the average of 100 intervals that fall within the same $|\sigma_c|$ bin. As shown in the right inset figure, the maximum value of kurtosis is decreasing with increasing $|\sigma_c|$, indicating that Alfv\'enicity is negatively correlated with intermittency. However, it has been shown that the Alfv\'enic character of the field fluctuations in the solar wind strongly decreases with radial distance \citep{chen_evolution_2020, shi_alfvenic_2021}. Therefore, to distinguish between the effects of radial evolution and decrease in $\sigma_c$, we show in Figure \ref{fig:sigma_c}b, the dependence of $K_{max}$ as a function of $|\sigma_c|$ and $\tau_{adv}$. It is  readily noticed, that on average, highly Alfv\'enic intervals, exhibit lower $K_{max}$ values. Nevertheless, it is evident that for a any $|\sigma_c|$ row in Figure \ref{fig:sigma_c}, an increasing trend is observed for $K_{max}$ at larger $\tau_{adv}$ values. The increase may be explained by the fact that there is still a mixing of parallel and perpendicular intervals outlined in Section \ref{subsec:thetavb}.





\section{Summary and conclusions}\label{sec:Summary}

This study has tried to address the following question: How do the statistical signatures of turbulence and intermittency evolve as the solar wind expands in the inner Heliosphere?  \par
As already discussed in Section \ref{sec:intro}, intermittency lies at the heart of MHD turbulence in the solar wind. Thus, an improved understanding of its radial evolution can offer insights into some of the major open problems in the field of space physics, including the origins of the fluctuations and coherent structures observed in the solar wind; the influence of local and global dynamics in the evolution of the higher-order statistics; and ultimately into fundamental questions, such as the generation, acceleration and adiabatic expansion of solar and stellar winds. For this purpose, we have analyzed high-resolution magnetic field and particle data from the first 11 orbits of the Parker Solar Probe mission, as well as Solar Orbiter observations, ranging from the vicinity of the Alfv\'en region ($R\approx 13.7 \ R_{\odot}$) out to 1 au ($R\approx 215 \ R_{\odot}$). 
Our study has been made possible by a variety of statistical tools, such as the Scale Dependent Kurtosis, the normalized scaling exponents of the Structure functions, and the PVI  method that enable us to exploit the property of PDFs of intermittency affected magnetic fluctuations to be increasingly flared out at progressively smaller scales.


The main findings of our study can be summarized as follows:

    \vspace{1em}
    (1) When methods utilizing higher-order moments are considered (e.g., $SDK$, $SF_{q}$), a strengthening of small-scale intermittency is observed with increasing advection time of the solar wind. Closer to the Sun,  fluctuations of spatial scale $\ell \approx 20-10^{2}d_i$, exhibit a monofractal-like but Super-Gaussian scaling that gradually evolves into multifractal as the solar wind expands into the interplanetary medium. Deeper in the inertial range, a multifractal scaling is observed that does not exhibit clear signs of radial evolution.
    \vspace{1em}

    (2) The PVI method provides a different perspective on the evolution of intermittency. For lag $\ell =20 d_i$, the fraction of the dataset occupied by coherent structures, $f_{PVI \geq 3}$, displays a very subtle upward radial trend, whereas a more obvious increase is observed for $f_{PVI \geq 6}$. At larger spatial scales $\ell =5\cdot10^{2}, \ 10^{3} d_i$, the opposite trend is observed as $f_{PVI \geq 6}$ is, within the error bars, independent of radial distance, while an increasing trend is observed for $f_{PVI \geq 3}$. It is important to note, however, that even though the trend remains positive at larger $\tau_{adv}$, the biggest gain is observed for $\tau_{adv} \lesssim 35$ hrs.
    
     \vspace{0.6em}

   In an effort to explain the disparity between SDK and PVI on the radial evolution of intermittency, the relationship between $f_{PVI \ge \theta}$ and the maximum values of SDK, $K_{max}$ was examined. We have shown that the fractional volume of events with $PVI \geq 6$ is strongly correlated with $K_{max}$. In light of this result, we can understand that methods relying on the estimate of higher-order moments as a measure of intermittency will mostly be affected by the extreme events that lie at the very tails of the distribution of increments. Such events are usually characterized by PVI values of the order of $PVI \gtrsim 6$ and constitute only a minor fraction $\lesssim 0.2 \%$ of the fluctuations observed in the solar wind. As a result, higher peaks in SDK may still be observed at larger $\tau_{adv}$ even though $f_{PVI\geq 3}$ stays constant as long as $f_{PVI \geq 6}$ radially increases. However, to fully characterize the radial evolution of intermittency, one has to also take into account the evolution of coherent structures with $PVI \geq 3$, as these structures have been shown to dissipate a considerable amount of magnetic energy in the solar wind \citep{osman_intermittency_2012}. In other words, a comprehensive analysis of the radial evolution of intermittency in the solar wind requires the use of lower-order moment-based methods, such as PVI.

     \vspace{0.3em}
   
    (3) CSS can both decay and reform due to local plasma dynamics during the expansion, with in situ generation being more efficient at the larger scales (Figure \ref{fig:PVI1}). Nevertheless, the existence of passively advected coherent structures of Solar origin cannot be ruled out. 
  
     \vspace{0.3em}

    An observation that warrants a brief discussion is the abrupt increase incoherent structures of $PVI \geq 3$ at the largest spatial scales in the vicinity of the Alfv\'en region. Recently \cite{Tenerani_2021} have analyzed PSP, Helios, and Ullyses data to show that the evolution of the occurrence rate of Switchbacks in the solar wind is scale-dependent as the fraction of longer-duration switchbacks increases with radial distance, whereas it decreases for shorter switchbacks. The PVI method is agnostic to the nature of the discontinuities, meaning that coherent structures may be identified by PVI as long as there are strong gradients in the magnetic field. As a result, several types of coherent structures such as current sheets, vortices, reconnection exhausts, and switchbacks may be identified with the PVI method. In this sense, one contributing factor to the increasing fraction of the dataset occupied by coherent structures might be the increasing trend of longer-duration switchbacks associated with increasing solar wind advection times. At the same time, several mechanisms, including stream-stream dynamic interactions, parametric decay instability of large amplitude Alfv\'en waves, \citep[e.g.,][]{Biskamp_muller_2000,2001NPGeo...8..159M,wan_oughton} might coexist simultaneously, resulting in the generation of several types of CSS. 

    \vspace{1em}
    (4) In agreement with earlier studies, we identify a strong anisotropy in intermittency with respect to the angle between the background magnetic field and the solar wind flow. Intermittency is weaker at $\Theta_{VB} \approx 0^{\circ}$ and is progressively strengthened at larger angles. More specifically, peaks in the SDK ($K_{max}$) are shifted upward, and an increase is observed in the fraction of the dataset occupied by coherent structures ($PVI \ge 3$) when intervals with increasingly larger $\Theta_{VB}$ angles are considered. The anisotropy is more pronounced at higher PVI thresholds but becomes weaker at progressively larger spatial scales ;
    

    \vspace{1em}
    (5) Even though at the smallest scales ($\ell =20 d_{i}$), the fraction of the dataset occupied by coherent structures ($PVI \ge 3$) radially decreases for intervals with fixed $\Theta_{VB}$, on average (i.e., averaging over all $\Theta_{VB}$ bins at a given $\tau_{adv}$ column in Figure \ref{fig:PVI-tadv-theta-vb-heatmap}) the fraction of measured coherent structures increases in the inner Heliosphere. This is because closer to the Sun, the solar wind flow is statistically (anti)parallel to the magnetic field (i.e., $\Theta_{VB} \approx 0^{\circ} (180^{\circ})$. However, due to the radial evolution of the Parker Spiral, the fraction of observed parallel intervals gradually decreases. Note that the changing fraction of parallel vs. perpendicular intervals is due to the limitations of the single-point measurements by PSP and the use of the Taylor hypothesis and not a reflection of the radial evolution of turbulence. Taking the $\Theta_{VB}$ anisotropy into account (see point (3) ), we can understand that the mixing of highly intermittent perpendicular and relatively less intermittent parallel intervals blur the averaged behavior of the radial evolution of intermittency. Due to the fact that the anisotropy is stronger at a higher $PVI$ threshold, the averaged fraction of events with $PVI \geq 6$ shows a more prominent positive radial trend with $\tau_{adv}$. This increase is also reflected in $K_{max}$, as already discussed in (2).

     \vspace{1em}
     
    (6) Solar wind of lower speed exhibits higher SDK peaks and is characterized by a higher fraction of $f_{PVI \ge 6}$ events. However, no statistically significant differences are observed in $f_{PVI \ge 3}$ as a function of solar wind speed. A strengthening of intermittency with respect to the advection time $\tau_{adv}$ is observed regardless of solar wind speed. (see Figures \ref{fig:VSW_intermit1}, \ref{fig:VSW_intermit2}, \ref{fig:VSW_intermit3})
    
    \vspace{1em}
   (7) A negative correlation is observed between the absolute value of the normalized cross-helicity and intermittency of the magnetic field. That is, Alfv\'enic intervals statistically display lower levels of intermittency as indicated by the maximum value of the $SDK$. (see Figure \ref{fig:sigma_c})


As already discussed in point 4, the mixing of parallel and perpendicular intervals in the inner Heliosphere will result in a subtle radial increase in intermittency. However, perpendicular intervals are expected to progressively dominate with increasing heliocentric distance due to the conservation of magnetic flux (i.e., Parker Spiral). It is, therefore natural to expect that when the fraction of parallel intervals becomes statistically insignificant, the decreasing trend in the fraction of the dataset occupied by coherent structures will become apparent. Several observational studies have indicated that intermittency is expected to become progressively weaker with increasing heliocentric distance beyond 1AU \citep{Parashar_2019, cuesta_intermittency_2022}. Based on our analysis, we propose that the decreasing trend in intermittency beyond 1AU can be attributed to the fact that the mixing effect is diminished due to the dominance of perpendicular intervals.

Finally, our results indicate that when it comes to analyzing the radial evolution of turbulence and intermittency, monitoring the changes in sampling direction is crucial. As the interplanetary magnetic field follows the Parker spiral, its angle with the spacecraft sampling direction will also vary as the distance from the Sun increases, which will then have an effect on measured turbulence characteristics. This effect needs to be disentangled from observations before the nature of the radial evolution of turbulence can be revealed. Previous studies using PSP data to analyze the radial evolution of intermittency have not taken this effect into consideration. However, our analysis indicates that obtaining such information is essential for understanding the more complex dynamics of the solar wind in the inner Heliosphere and can facilitate improvements to simulations of the solar wind \citep[see also,][]{ 2020ApJ...898..113Z,2021ApJ...923...89C, https://doi.org/10.48550/arxiv.2205.00526}.
Our results will further the understanding of how CSS are generated and transported in the solar wind and will guide the development of future solar wind turbulence
models.


\vspace*{0.1cm}

\begin{acknowledgements}

We would like to thank Professor Tim Horbury and the the magnetometer instrument team of the Solar Orbiter mission.
 This research was funded in part by the FIELDS experiment on the Parker Solar Probe spacecraft, designed and developed under NASA contract NNN06AA01C; the NASA Parker Solar Probe Observatory Scientist grant NNX15AF34G and the  HERMES DRIVE NASA Science Center grant No. 80NSSC20K0604.  \par Solar Orbiter is a space mission of international collaboration between ESA and NASA, operated by ESA. Solar Orbiter Solar Wind Analyser (SWA) data are derived from scientific sensors which have been designed and created, and are operated under funding provided in numerous contracts from the UK Space Agency (UKSA), the UK Science and Technology Facilities Council (STFC), the Agenzia Spaziale Italiana (ASI), the Centre National d’Etudes Spatiales (CNES, France), the Centre National de la Recherche Scientifique (CNRS, France), the Czech contribution to the ESA PRODEX programme and NASA. Solar Orbiter SWA work at UCL/MSSL is currently funded under STFC grants ST/T001356/1 and ST/S000240/1.

\end{acknowledgements}

\section{Appendix}\label{sec:Appendix}



\begin{figure}[htb!]
\centering
\includegraphics[width=0.4\textwidth]{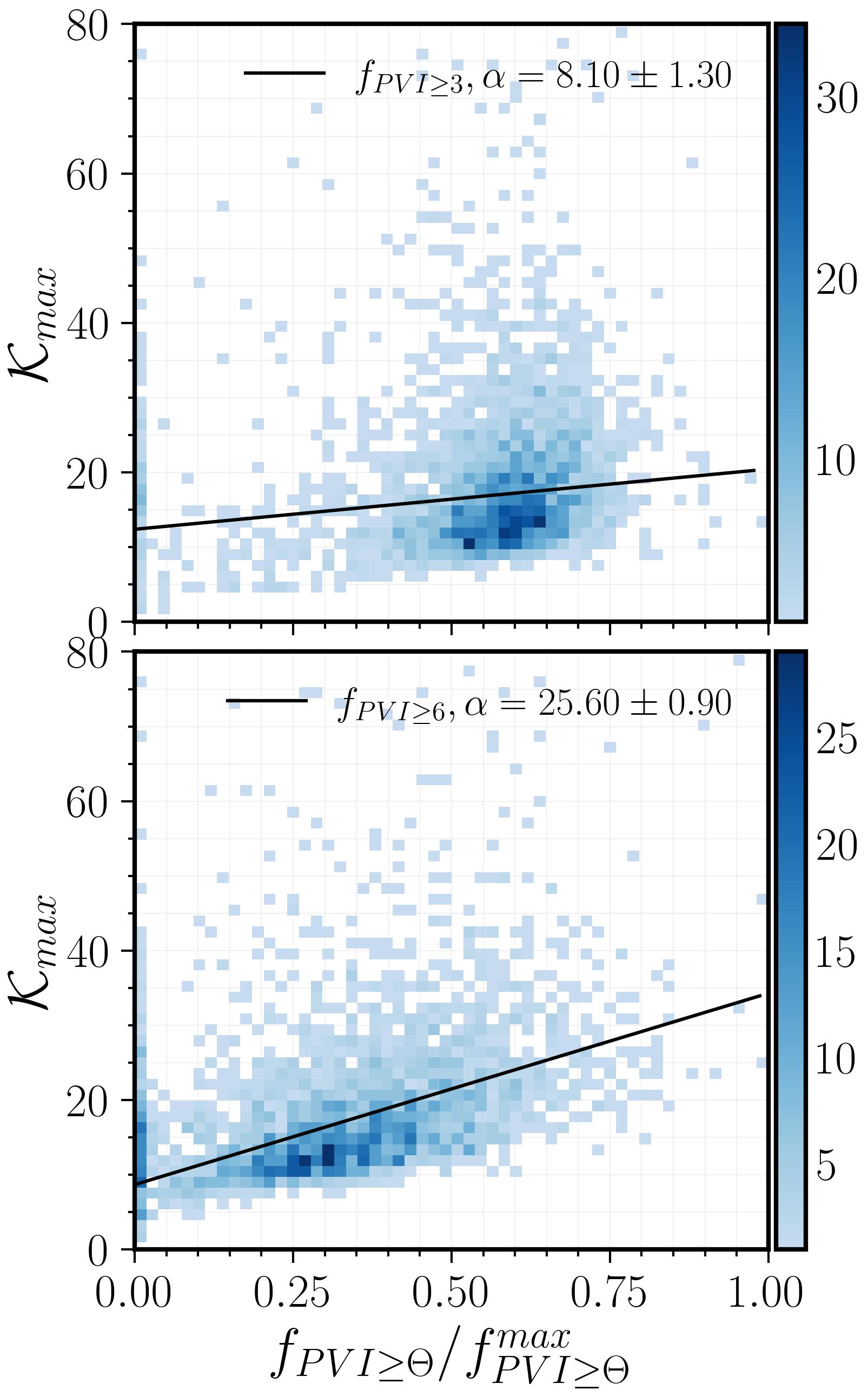}

\caption{The maximum of SDK, $K_{max}$, is plotted as a function of the fraction of PVI events that exceed a threshold $\theta$, $f_{PVI \ge \theta}$ for $\theta =3,\  6$, shown in the top and bottom pannel respectively. For a more direct comparison of the two curves $f_{PVI \ge \theta}$ was normalized to the maximum value $f^{max}_{PVI \ge \theta}$ estimated for the respective threshold.}
\label{fig:PVI_kurt_appendix}
\end{figure}

In this Appendix, we investigate the relationship between the maximum of SDK, $K_{max}$, and the fraction of PVI events that exceed a threshold $\theta$, $f_{PVI \ge \theta}$ for $\theta =3,\  6$. For this reason, a new set of runs was initiated. The maximum value of kurtosis as well as $f_{PVI \geq 3}$, and $f_{PVI \geq 6}$, was estimated for intervals of duration $d= 5 hr$. In Figure \ref{fig:PVI_kurt_appendix} the maximum value of Kurtosis $K_{max}$ is plotted  against $f_{PVI \geq 3}$ (blue line), and $f_{PVI \geq 6}$ (red line). As already shown in Figure \ref{fig:PVI1}, $f_{PVI \geq \theta}$ for $\theta =3, 6$  obtain values that are separated by at least one order of magnitude. Consequently, for a more direct comparison, $f_{PVI \geq \theta}$ values were normalized by their corresponding maximum observed value. Though, in both cases, the underlying scatter plot does not exhibit a clear linear relationship (especially for $f_{PVI \geq 3}$), a linear fit was applied to provide a rough estimate for the relationship between $K_{max}$ and $f_{PVI \geq \theta}$. From this analysis, we can see that the slope for $f_{PVI \geq 6}$ is significantly steeper than the one corresponding to $f_{PVI \geq 3}$. Similar ratios for the slopes were obtained when the quantities were normalized by the mean or their standard deviation.

\bibliography{sample631}{}
\bibliographystyle{aasjournal}



\end{CJK*}
\end{document}